\documentclass[twocolumn, superscriptaddress, amsmath, amssymb, pre]{revtex4-1}
\usepackage[utf8]{inputenc}
\usepackage{comment}
\usepackage[T1]{fontenc}
 \usepackage{mathtools}
\usepackage{tikz}
\usetikzlibrary{decorations.pathmorphing}
\usetikzlibrary{decorations.markings}
\usetikzlibrary{arrows,shapes,snakes,automata,backgrounds,petri}
\usetikzlibrary{calc}
\usetikzlibrary{patterns}
\usetikzlibrary{decorations.text}
\tikzset{
xxtsubstrate/.style={decorate, 
line width=1pt,
draw=olive, 
decoration=snake, 
segment amplitude=0.75mm, 
line after snake=0.25mm,
line before snake=0.25mm
},
tsubstrate/.style={decorate, 
line width=1pt,
draw=olive, 
decoration=snake, 
segment amplitude=0.5mm, 
segment length=5pt,
segment amplitude=0.2mm, 
line after snake=1mm,
line before snake=1mm
},
Bsubstrate/.style={decorate, 
line width=1pt,
draw=olive, 
decoration=snake,
segment length=5pt,
segment aspect=0,
segment amplitude=0.5mm, 
line after snake=0mm,
line before snake=0mm
},
substrate/.style={decorate, 
very thick,
draw=orange, 
decoration=snake, 
segment length=5pt,
segment amplitude=0.5mm, 
line after snake=0.5mm,
line before snake=0.5mm
},
activity/.style={very thick,draw=red,postaction={decorate},
decoration={markings,mark=at position .5 with
{\arrow[draw=red]{>}}}},
tactivity/.style={thick,draw=red,postaction={decorate},
decoration={markings,mark=at position .5 with
{\arrow[draw=red]{>}}}},
tEPSactivity/.style={thick,draw=red,postaction={decorate},
decoration={markings,mark=at position .55 with
{\arrow[draw=red]{>}}}},
tAactivity/.style={thick,draw=red},
Aactivity/.style={very thick,draw=red},
Cactivity/.style={very thick,draw=blue,decorate,decoration={snake}},
Bactivity/.style={very thick,draw=blue,densely dashed},
tSactivity/.style={thick,draw=red,postaction={decorate},
decoration={markings,mark=at position .7 with
{\arrow[draw=red]{>}}}},
Sactivity/.style={very thick,draw=red,postaction={decorate},
decoration={markings,mark=at position .7 with
{\arrow[draw=red]{>}}}}
}

\usepackage[colorlinks=true,linkcolor=blue, urlcolor = black, citecolor = blue, anchorcolor = blue]{hyperref}

\usepackage{graphicx}
\usepackage[tight]{subfigure}
\subfiguretopcaptrue
\usepackage{multirow}
\usepackage{subeqnarray}
\usepackage{wasysym}
\usepackage{eufrak}

\newcommand{\ave}[1]{\left\langle #1 \right\rangle}

\newcommand{\plaind}{\mathrm{d}}

\newcommand{\tildephi}{\widetilde\phi}

\newcommand{\Eref}[1]{Eq.~(\ref{eq:#1})}

\newcommand{\eref}[1]{(\ref{eq:#1})}

\newcommand{\elabel}[1]{\label{eq:#1}}
\newcommand{\deltabar}{\delta\mkern-8mu\mathchar'26}
\newcommand{\imag}{\mathring{\imath}}
\newcommand{\dbar}{\plaind\mkern-6mu\mathchar'26}
\newcommand{\dint}[1]{\mathchoice{\!\plaind#1\,}{\!\plaind#1\,}{\!\plaind#1\,}{\!\plaind#1\,}}

\newcommand{\dintbar}[1]{\mathchoice{\!\dbar#1\,}{\!\dbar#1\,}{\!\dbar#1\,}{\!\dbar#1\,}}

\newcommand{\Exp}[1]{\operatorname{exp}\!\left(#1\right)}
\renewcommand{\exp}[1]{\mathchoice{e^{#1}}{\operatorname{exp}\!\left(#1\right)}{\operatorname{exp}\!\left(#1\right)}{\operatorname{exp}\!\left(#1\right)}}
\newcommand{\corresponding}{\hat{=}}

\newcommand{\gcomment}[1]{\textcolor{orange}{[GP: #1]}}
\newcommand{\jcomment}[1]{\textcolor{teal}{[JP: #1]}}

\newcommand{\mA}{\mathcal{A}}
\newcommand{\mD}{\mathcal{D}}

\newcommand{\mO}{\mathcal{O}}
\newcommand{\Num}{N}

\usepackage{xspace}
\newcommand{\latin}[1]{{\it #1}}
\newcommand{\ie}{\latin{i.e.}\@\xspace}

\newcommand{\cf}{\latin{cf.}\@\xspace}

\newcommand{\Erefs}[1]{Eqs.~(\ref{eq:#1})}

\newcommand{\mean}[1]{\mathbb{E}\left[#1\right]}
\newcommand{\meanT}[1]{\mathbb{E}_\tau\left[#1\right]}
\newcommand{\meanY}[1]{\mathbb{E}_y\left[#1\right]}
\newcommand{\muA}{\mu_{A}}
\newcommand{\muB}{\mu_{B}}

\allowdisplaybreaks

\begin{document}

\title{
Noise can lead to exponential epidemic spreading despite $R_0$ below one
}

\author{Johannes Pausch}
\affiliation{Department of Applied Mathematics and Theoretical Physics, University of Cambridge, Cambridge CB3 0WA, United Kingdom}
\affiliation{St.~Catharine's College, Cambridge CB2 1RL, United Kingdom}
\author{Rosalba Garcia-Millan}
\affiliation{Department of Applied Mathematics and Theoretical Physics, University of Cambridge, Cambridge CB3 0WA, United Kingdom}
\author{Gunnar Pruessner}
\affiliation{Department of Mathematics, Imperial College London, London SW7 2AZ, United Kingdom}
\date{\today}

\begin{abstract}
    Branching processes are widely used to model evolutionary and population dynamics as well as the spread of infectious diseases. To characterize the dynamics of their growth or spread, the basic reproduction number $R_0$ has received considerable attention. In the context of infectious diseases, it is usually defined as the expected number of secondary cases produced by an infectious case
     in a completely susceptible population. Typically $R_0>1$ indicates that an outbreak is expected to continue and to grow exponentially, while $R_0<1$ usually indicates that an outbreak is expected to terminate after some time. 
    
    In this work, we show that fluctuations of the dynamics in time can lead to a continuation of outbreaks even when the expected number of secondary cases from a single case is below $1$. Such fluctuations are usually neglected in modelling of infectious diseases by a set of ordinary differential equations, such as the classic SIR model. We showcase three examples: 1) extinction following an Ornstein-Uhlenbeck process, 2) extinction switching randomly between two values and 3) mixing of two populations with different $R_0$ values. We corroborate our analytical findings with computer simulations.
\end{abstract}

\maketitle
\newcommand{\eon}{\epsilon_{\text{on}}}
\newcommand{\eoff}{\epsilon_{\text{off}}}
\newcommand{\mon}{\mu_{\text{on}}}
\newcommand{\moff}{\mu_{\text{off}}}

\section{Introduction}
A branching process \cite{Harris:1963,AthreyaNey:1972} is a stochastic process in which individuals randomly create copies of themselves or become extinct. This model has a wide range of applications and individuals might represent offspring \cite{Watson:1875,Nowak:2006,Bacaer:2011}, particles \cite{Pazsit2007,Williams2013,Garcia-Millan2018}, active neurons \cite{Wilting2018,Zierenberg2018,Pausch2020,Zierenberg2020,Pausch2021}, or infected individuals \cite{Farrington1999,Farrington2003,Wilting2018,Corral:2021} among others \cite{kimmel2002}.
At the centre of many investigations of branching processes is the statistics of spells of activity, which are called avalanches \cite{Pazsit2007,Williams2013,Garcia-Millan2018,Wilting2018,Zierenberg2018,Pausch2020,Zierenberg2020,Pausch2021} in most applications or outbreaks in the context of infectious diseases. Avalanches, or outbreaks, are defined as the activity in the branching process which is initiated by a single individual and lasts until all subsequent individuals have become extinct. In the context of particles, an avalanche starts with a single particle and ends when the system is empty. In the context of infectious diseases, an outbreak starts with a single infected individual and ends when no infected individual is left (here we 
use the words \emph{infected} and \emph{infectious} without
distinction). 

A branching process is subcritical if the 
expected size of an outbreak \emph{decays} exponentially.
It is called supercritical if the expected size of the
outbreak \emph{grows} exponentially. 
Otherwise, if the size approaches a constant value in time, the 
branching process is said to be critical. 
We will use this criterion, asymptotically constant expected outbreak size, as the definition of the critical point throughout this work.
The critical point generally divides a parameter region resulting in asymptotically exponential growth from one resulting in asymptotically exponential decline.
For particle avalanches as well as for outbreaks of infectious diseases, it is of great interest to identify simple parameters that indicate whether the process is supercritical or not. One of those parameters is the basic reproduction number $R_0$.

The basic reproduction number $R_0$ is defined as the expected number of secondary individuals that are created from a single individual \cite{Dietz1993,Heesterbeek1996,Heesterbeek2002,LiJing2011,Delamater2019}. More explicitly in a branching process, an existing individual waits until a branching event occurs. 
In many models, the waiting time between branching events is fixed \cite{Harris:1963,CorralETAL:2018branchings,CorralETAL:2016}, however in the present work, we consider only exponentially distributed waiting times \cite{BordeuETAL:2019,Garcia-Millan2018,Pausch2020}. 
At a branching event, an individual is replaced by its offspring, which is a random number $K\in\mathbb{N}_0$ of individuals.
The case $K=0$ corresponds to the extinction of the parent individual. 
In general, the offspring distribution can be defined by specifying the branching probabilities $p_0$, $p_1$, $p_2$, \ldots$\in[0,1]$ such that
\begin{align}
    P(K=k)=p_k\qquad\text{with}\qquad\sum\limits_{k=0}^\infty p_k=1,
    \elabel{def_p_k}
\end{align}
\ie~the probability $P(K=k)$ that an individual has $K=k$ offspring equals $p_k$. 
Although our analytical results hold for any distribution of $K$, in our simulations we use only the binary offspring distribution with $K\in\{0,2\}$.
All individuals are independent and there is no bound to the number of individuals in the system, which can be interpreted as an unbounded population of susceptible individuals.  In this setup, the basic reproduction number $R_0$ is defined as the expected number of offspring,
\begin{align}
    R_0=
    \sum_{k=0}^\infty k p_k =
    \mean{K}.
    \label{def_R0}
\end{align}
In particular, $R_0$ is dimensionless and does not indicate how quickly or slowly an avalanche/outbreak evolves. 

There are many difficulties in deriving $R_0$ from data \cite{LiJing2011,Diekmann1990,vandenBosch2008} and researchers have defined several similar quantities related to $R_0$ \cite{Anderson1992,CaoETAL:2020}.
In addition, more detailed models of infectious diseases take other characteristics such as age, immunity, behaviour or the evolution of the disease itself into account, which make the definition of $R_0$ more difficult \cite{Diekmann1990,Anderson1992,vandenBosch2008,LiJing2011,Ridenhour2014}. Rather than including more detailed aspects into our models, we restrict ourselves to the basic model outlined above and keep the discussion at the level of stochastic processes. For example, we assume that infected 
individuals are also infectious.

In many real-world occurrences of branching processes, the environment and the process itself are imperfect in the sense that they fluctuate in time, for example because individuals and the environment change the conditions for disease transmission \cite{ArielLouzoun:2021}.
Such fluctuations will be affecting the branching process over time and are not easily dealt with analytically. 
Of particular importance are fluctuations that affect the population as a whole.
As our calculations show, basic approximations of branching processes with noise can be misleading by predicting subcritical dynamics where a more detailed calculation reveals supercritical behaviour. It is the main aim of this article to highlight such, often counter-intuitive, phenomena, which are easily missed by traditional modelling of epidemics, which draw on a coarse-grained set of equations, such as the classic SIR model \cite{VasiliauskaiteAntulov-FantulinHelbing:2021}.

The article is organized as follows: In Sec.~\ref{Sec:Birth-Death-Model}, the branching model without noise is presented as a Master Equation. It forms the basis of the models with noise that follow. 
In Sec.~\ref{Sec:LinearOU}, we introduce a branching process coupled to an Ornstein-Uhlenbeck process. Although a mean-field approximation predicts a critical point at $R_0=1$, our detailed analysis supported by simulations reveal a shift of the critical $R_0$ to values smaller than $1$. 
In Sec.~\ref{Sec:TelegraphicNoise}, the branching process is coupled to a stochastic process called telegraphic noise, which implements a random switch between two different extinction rates of the branching process. In this system, it is much less obvious what a suitable definition of $R_0$ would be. Neither of the two $R_0$ values associated with the two extinction rates, nor a simple weighted average of them predict $R_0=1$ to be the critical point. An exact calculation, reveals that the critical $R_0$ is smaller than $1$. 
In Sec.~\ref{Sec:Coupled-Birth-Death-Process}, two branching processes with two different $R_0$ are coupled. We show that neither of the two $R_0$ nor a linear combination of them correctly predict the critical point of the system. We conclude in Sec.~\ref{Sec:Conclusion}. The detailed analytical results are based on field-theoretic approaches which are presented in the appendix.



\begin{figure}
    \centering
    \includegraphics[width=0.5\textwidth]{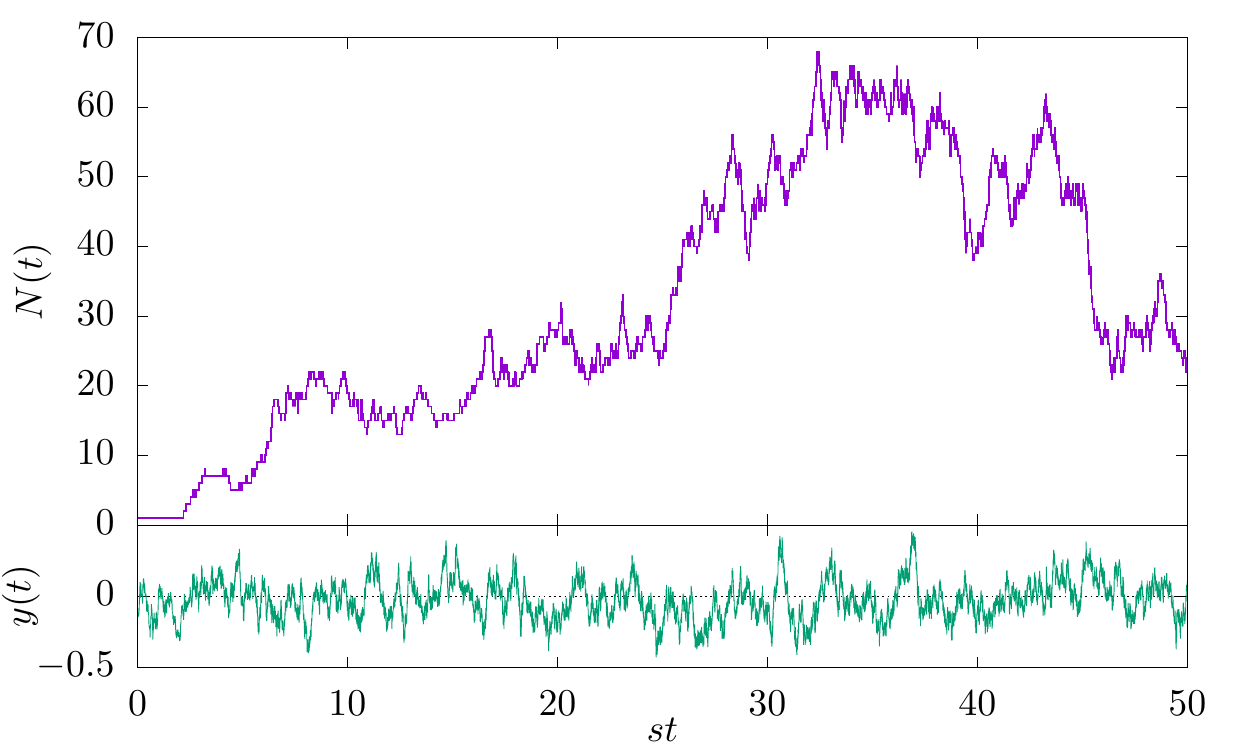}
    \caption{Trajectory of a
    branching processes (top) whose extinction rate is coupled to an Ornstein-Uhlenbeck process (bottom),
    see Sec.~\ref{Sec:LinearOU}.
    Parameters for this simulation are
    $s=1$, $R_0=1$, $\lambda=0.18$, $\beta=3.5$ and $D=2$.}
    \label{fig:traj_BP_OU}
\end{figure}

\begin{figure}
    \centering
    \includegraphics[width=0.5\textwidth]{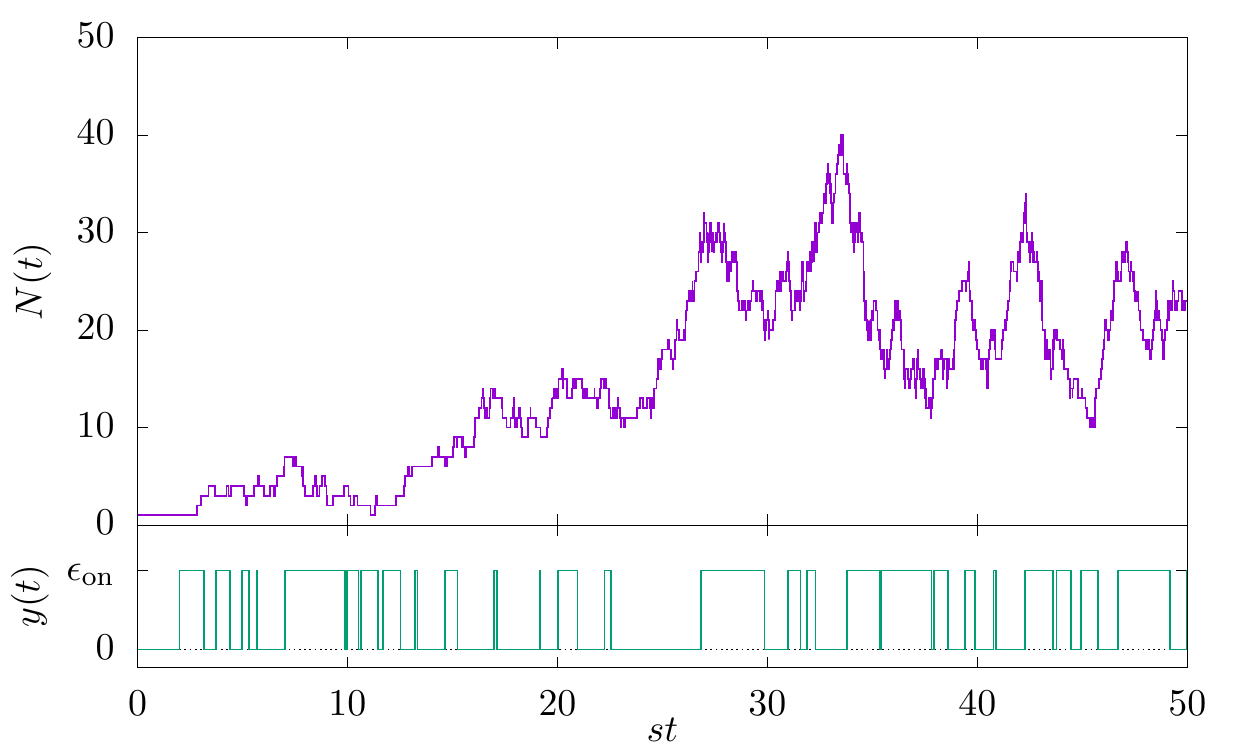}
    \caption{Trajectory of a
    branching process (top) whose extinction rate is coupled to a Telegraphic noise (bottom),
    see Sec.~\ref{Sec:TelegraphicNoise}.
    Parameters for this simulation are
    $s=1$, $R_0=1.2$, $\mon=\moff=1$, $\eon=0.444$ and $\eoff=0.000$.}
    \label{fig:traj_BP_T}
\end{figure}

\begin{figure}
    \centering
    \includegraphics[width=0.5\textwidth]{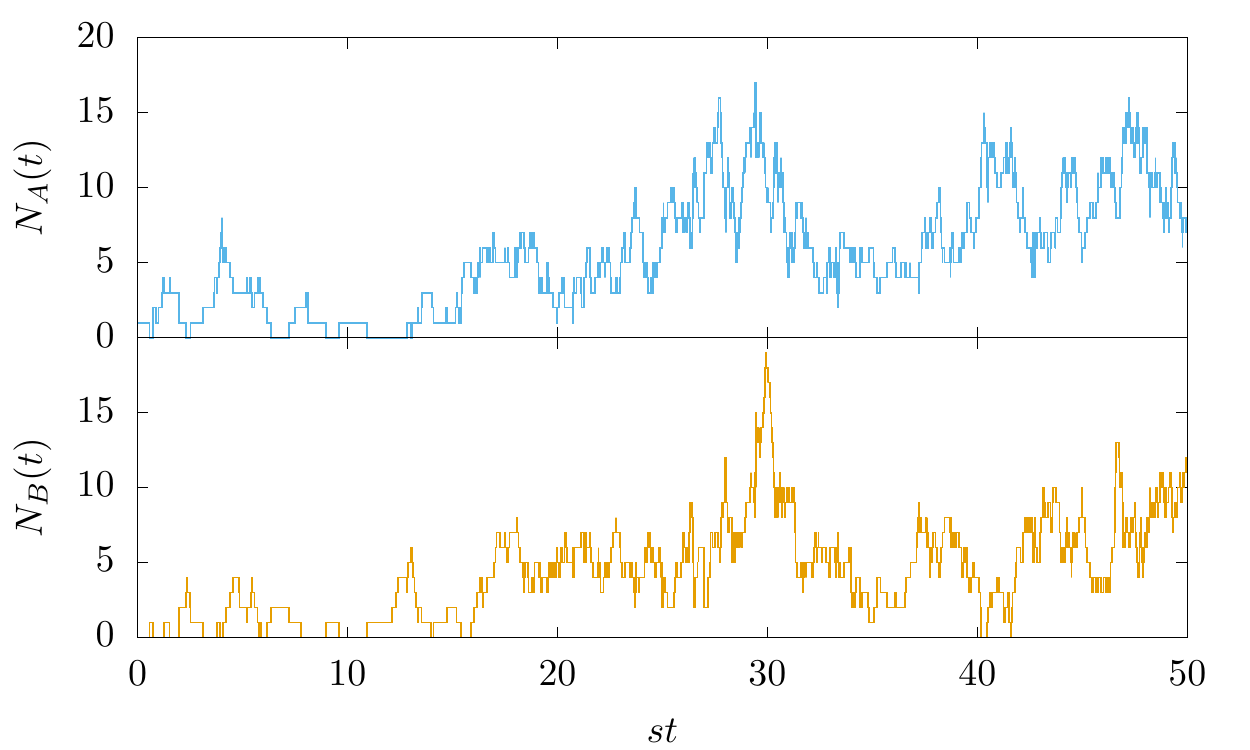}
    \caption{Trajectories of two
    coupled branching process with $s=1$, 
    basic reproduction numbers
    $R_{0A}=1.2$, $R_{0B}=0.75$ and transmutation rates
    $\muA=\muB=1$, see Sec.~\ref{Sec:Coupled-Birth-Death-Process}.}
    \label{fig:traj_coupledBP}
\end{figure}

\section{Basic Branching Process}
\label{Sec:Birth-Death-Model}
At the center of our models is the Master Equation for a branching process \cite{Garcia-Millan2018}. In order to use a consistent language, we refer to \textit{individuals} being \textit{created} or becoming infected in branching events. They are then \textit{present} in the \textit{population}. We will say that they \textit{disappear} from the population when an extinction occurs. We hope that the reader can translate this language to their application by replacing individual with particle, signal, ... and population with system, neural network, ... according to their requirements.

Let the population size  $N(t)$ denote the number of individuals present at time $t$, with initial condition $N(0)=1$, and $P(N(t)=n)$ the probability that there are $n$ individuals present at time $t$. Then, the probabilistic dynamics of the basic branching process are described by
the master equation
\cite{Garcia-Millan2018}
\begin{align}\label{eq_master_branching}
    {\partial_t}
    P(N;t)=&\,s\sum\limits_{k=0}^\infty p_k(N-k+1)P(N-k+1;t)\notag\\&\,-s N P(N;t),
\end{align}
where $p_k$ is the offspring distribution and $s$ is the overall event rate. In particular, there are two probabilistic components: 1) $s$ is the parameter for an exponentially distributed waiting time until a branching event occurs, and 2) at a branching event of an individual, the offspring distribution $p_k$ determines by how many individuals it is \textit{replaced}, \ie~the original individual does not continue to exist alongside the $k$ new individuals. Alternatively, one could say that $k-1$ new individuals are created while the original individual continues to be present.


The independence of the processes implies that the product $sp_k$ equals the event rate for an individual to be replaced by $k$ other individuals. In the case $k=0$,
we denote by $\epsilon:= sp_0$ the extinction rate, \ie~the rate of the exponential distribution that determines the time when an individual disappears from the population without producing any offspring. In the present work, we consider branching dynamics for which $\epsilon$ fluctuates in time. To clarify the r{\^o}le of the extinction rate $\epsilon$, we rewrite Eq.~\eqref{eq_master_branching} as
\begin{align}\label{eq_master_branching_with_eps}
    &{\partial_t} P(N;t)\notag\\
    =&\,s\sum\limits_{k=2}^\infty p_k\Big\{(N-k+1)P(N-k+1;t)
    - N P(N;t)\Big\} \notag\\
    &\,+\epsilon 
    \Big\{
    (N+1) P(N+1;t)
    -N P(N;t)
    \Big\} \,.
\end{align}


In our approach, there is no modelling of a healthy population and there is no saturation of a population with infected individuals. In particular, there is no upper bound to the number of infected individuals. 
This is because we want to identify the effect of the noise on the extinction rate without getting lost in the too many details and model parameters.

The time-homogeneous branching process described by the Master Equation~\eqref{eq_master_branching} has been studied before \cite{Harris:1963, Garcia-Millan2018}.
The temporal evolution of the expected number of infected individuals, 
\begin{equation}
    \mean{N(t)} = \sum_{N=0}^\infty N P(N;t) \,,
\end{equation}
follows from \eqref{eq_master_branching}, \begin{align}\label{eq_first_moment_branching}
    {\partial_t}\mean{N(t)}=s(R_0-1)\mean{N(t)},
\end{align}
with $R_0$ as defined in Eq.~\eqref{def_R0}. 
The solution $\mean{N(t)}=\exp{s(R_0-1)t}$ illustrates the 
important role of the basic reproduction number $R_0$: if $R_0<1$, the expected number of individuals decreases over time (subcritical case), while it increases exponentially if $R_0>1$ (supercritical case). The case $R_0=1$ is called the critical point as the expected number of individuals
stays constant in time. 

The model described by Eq.~\eqref{eq_master_branching} does not rely on the law of large numbers, \ie~it is valid for small numbers and even for one individual. The equation also allows deriving non-trivial dynamics for higher moments. For example, the equation that governs the variance $\mathbb{V}[N(t)]$ of the number of infected individuals is\begin{align}\label{eq_variance_branching}
    {\partial_t}\mathbb{V}[N(t)]=2s(R_0-1)\mathbb{V}[N(t)]+s\mean{(K-1)^2}\mean{N(t)}.
\end{align}
Here the appearance of $\mean{(K-1)^2}$ implies that the second moment of the offspring distribution affects the dynamics too -- not just $R_0$.

The model in Eq.~\eqref{eq_master_branching}, and therefore also Eqs.~\eqref{eq_master_branching_with_eps},
\eqref{eq_first_moment_branching}
and \eqref{eq_variance_branching}, assumes a static setup, \ie~branching always happens with the same rates and the same offspring distribution.
However, this assumption
 may be unrealistic for many applications. What if event rates and offspring distributions fluctuate in time? This is the topic of the next two sections, after which we also consider the case where two populations with different basic reproduction numbers $R_0$ interact. 

\section{Branching coupled to an Ornstein-Uhlenbeck process}
\label{Sec:LinearOU}
As a first example of noisy branching processes, we couple the extinction rate $\epsilon$ in Eq.~\eqref{eq_master_branching_with_eps} 
to an Ornstein-Uhlenbeck (OU) process
\begin{align}\label{epsilon_OU}
\epsilon(t)= sp_0+\lambda y(t),
\end{align}where the rate $y(t)$ is governed by
\begin{align}\label{eq_OU_def}
    {\partial_t} y(t) =-\beta y(t)+\eta(t).
\end{align}
Here, $\eta(t)$ is a Gaussian white noise with mean $\mean{\eta(t)}=0$ and correlator $\mean{\eta(t)\eta(t')}=2D\delta(t-t')$. The dimensionless parameter $\lambda$ is the coupling strength and $\beta$ is the return rate. 
The persistence time $\beta^{-1}$ induces temporal correlations, or
a memory, in the noise $y(t)$ in a similar way to active fluctuations
in the motion of active Ornstein-Uhlenbeck particles \cite{DabelowETAL:2019,WalterPruessnerSalbreux:2021}, \Eref{OU-correlation}.
In principle, $\epsilon(t)$ may become negative, which would render the process ill-defined. In numerical simulations, we can guard against that by monitoring the value of $\epsilon$ and replacing it by $0$ whenever it becomes negative. For the parameters considered below, this is exceedingly rare, affecting a single realisation in well over $10^{10}$. An example trajectory is shown in Fig.~\ref{fig:traj_BP_OU}.


Eq.~\eqref{eq_OU_def} implies that the steady state distribution of $y$ is the Gaussian distribution 
\begin{equation}\label{eq_steady_state_OU}
    \lim_{t\to\infty}P(y;t) = \sqrt{\frac{\beta}{2\pi D}} \exp{-\frac{\beta y^2}{2D}}\,
\end{equation}
and therefore the time average equals $\meanT{y}=0$. The \emph{noisy branching process} Eq.~(\ref{eq_master_branching_with_eps}) with Eq.~(\ref{epsilon_OU}) is described by the following master equation:
\begin{align}\elabel{eq_master_branching_OU}
    {\partial_t}P(N;t)=&\,s\sum\limits_{k=0}^\infty p_k(N-k+1)P(N-k+1;t)\notag\\
    &+\lambda y(N+1)P(N+1;t) \notag\\
    & -(s+\lambda y) NP(N;t) \,.
\end{align}
The additional contribution $y(t)$ to the extinction rate $\epsilon(t)$ is affecting all individuals equally, so that rather than being reduced by the law of large numbers, the effect of $y(t)$ grows linearly with the population size.

What is the effect of this noise on the dynamics of the branching process? A mean-field approximation predicts that this perturbation does not have any impact, because in the mean-field approach all occurrences of $y(t)$ are replaced by its mean $\meanT{y}=0$ and therefore the mean-field expected number of secondary cases equals $R_0$, Eq.~\eqref{def_R0}. In other words, mean-field theory predicts that $R_0=\sum k p_k=1$ results in a critical process.

However, closer inspection reveals that with OU noise, the offspring distribution $P(K=k)=p_k$, Eq.~\eref{def_p_k}, has effectively become time-dependent. To see this, we regard the branching process as a collection of simultaneous, independent, exponentially-distributed waiting processes --- one process for each $K\in\mathbb{N}_0$. Because they are independent, the waiting time until the first of these events occurs is exponentially distributed with rate 
$s+\lambda y(t)$. 
Which of the processes actually occurs first can be answered probabilistically by calculating the ratio of the rate of that process divided by the sum of the rates of all simultaneous processes: \begin{subequations}\begin{align}
    P(K=k)=&\,\frac{sp_k}{s+\lambda y}&&\text{ for }k\neq0,\\
    P(K=0)=&\,\frac{sp_0+\lambda y}{s+\lambda y}&&\text{ for }k=0,
\end{align}\end{subequations}
which satisfies normalisation, $\sum_{k\geq0}P(K=k)=1$, and
describes the effective offspring distribution given a rate $y=y(t)$. Hence, the time-dependent expected number of secondary cases equals 
\begin{align}\label{eq_time_dependent_R_OU}
    \mean{K(t)}=\frac{s}{s+\lambda y(t)}R_0,
\end{align}
with $R_0=\sum k p_k$, Eq.~\eqref{def_R0}.
The effect of $y(t)$ in Eq.~\eqref{eq_time_dependent_R_OU} is not symmetric about $0$, as can be seen by expanding $s/(s+\lambda y)=1-(\lambda y/s) + (\lambda y/s)^2 - \ldots$
Taking the expectation over $y$, suggests $\mean{K(t)}=R_0(1+(\lambda/s)^2 \mean{y^2} + \ldots)$.
As shown in Appendix~\ref{app_dawson},
the average $\meanY{\mean{K(t)}}$ of $\mean{K(t)}$ over the stationary distribution of $y$, Eq.~\eqref{eq_steady_state_OU}, effectively the time-average of $\mean{K(t)}$, can be calculated in closed form
\begin{align}\elabel{eq_R_0_OU}
    \meanY{\mean{K(t)}}=\frac{s}{\lambda}\sqrt{\frac{2\beta}{D}}{{\mD_+}}\!\left(\frac{s}{\lambda}\sqrt{\frac{\beta}{2D}}\right)R_0\, ,
\end{align}
where $\mD_+$ denotes the Dawson function, defined in \eref{Dawson_def}. 



\begin{figure*}
\includegraphics[width=\textwidth]{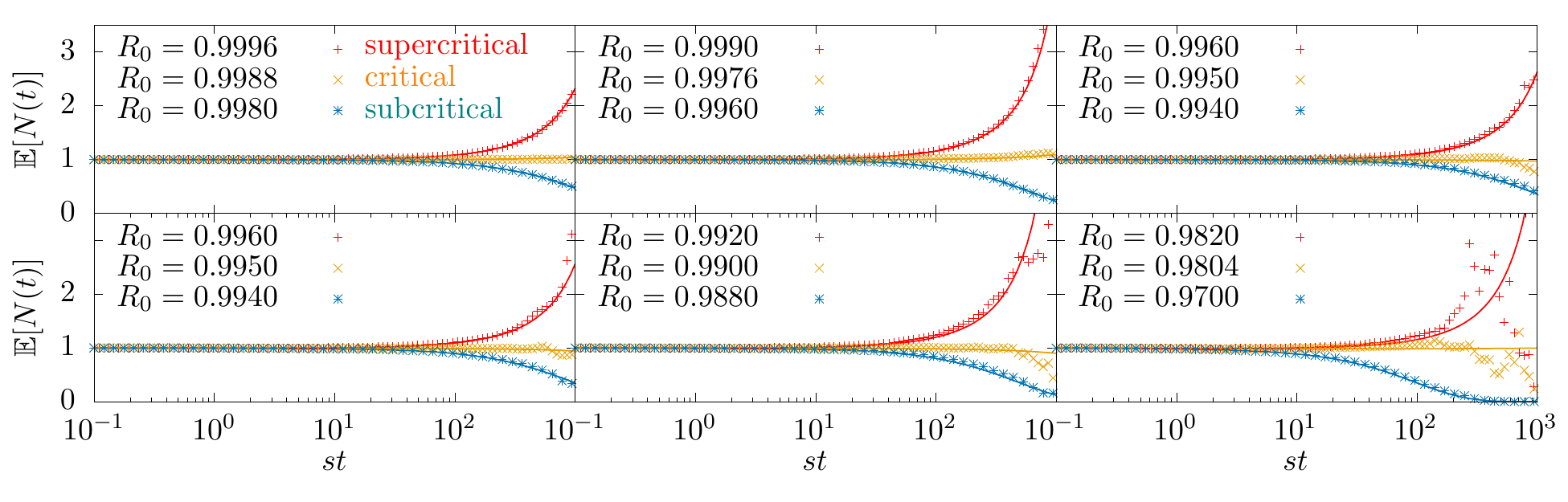}
\caption{\label{sample_OU_trajectories}
Expected number of infected individuals $\mean{N(t)}$ of a branching process
driven by an Ornstein-Uhlenbeck noise
estimated from numerical simulations (symbols) based on $10^7$ iterations.
The parameters of the noise in each panel are as follows:
$\lambda = 0.05$ (top row);
$\lambda =0.1$ (bottom row); 
$\beta=0.5$, $D=0.125$ (left column);
$\beta=1$, $D=1$ (middle column);
$\beta = 1$, $D=2$ (right column).
For each set of noise parameters, we can choose different values
of $R_0$ such that the population has 
supercritical (red symbols, black lines),
critical (orange symbols, grey lines),
and subcritical dynamics (blue symbols, pink lines).
Our numerical results are in agreement with \Eref{DysonSolution}.
In particular, we find values $R_0<1$ such that the population displays 
supercritical behaviour due to the external noise.
}
\end{figure*}


Demanding the expected offspring number 
$\meanY{\mean{K(t)}}$
to be unity at the critical point, produces
\begin{align}\elabel{R0_Dawson_expanded}
    R_0 = 1 
   - \frac{D \lambda^2}{\beta s^2}
   - 2 \left(
   \frac{D \lambda^2}{\beta s^2}
   \right)^2
   + \mathcal{O}\left(\left(\frac{D\lambda^2}{\beta s^2}\right)^3\right) \ ,
\end{align}
which generally is less than unity, as can also be gleaned from $R_0=1-(\lambda/s)^2 D/\beta + \ldots$ using the expansion discussed after Eq.~\eqref{eq_time_dependent_R_OU} and $\mean{y^2}=D/\beta$ from Eq.~\eqref{eq_steady_state_OU}. 

However, \eref{R0_Dawson_expanded} needs further scrutiny as it is based on the wrong assumption, as we will explain now.
The process is at the critical point when the average number of offspring spawned per reproductive event is unity. However, 
$\meanY{\mean{K(t)}}$ in \eref{R0_Dawson_expanded} is an average over the stationary distribution of $y$, 
assuming that the same number of reproductive events take place at any such value. 
Because of the temporal correlations in $y(t)$, however, episodes of high extinction typically occur when the population size is small anyway. As a result, fewer reproductive events are affected by high extinction rates than by low extinction rates.





Using a Doi-Peliti field theory, which is derived and explained in Appendix~\ref{sec:appendix-linear-OU}, we can calculate the expected population size directly, producing the final result
\begin{align}
\mean{\Num(t)}&= \exp{-\left(s(1-R_0)+\frac{\beta}{2}\right)t}
\left(\cosh\left(\frac{\beta t}{2}\sqrt{1+\frac{4D\lambda^2}{\beta^3}}\right)\right.
\nonumber\\&
\left.+\frac{1}{\sqrt{1+\frac{4D\lambda^2}{\beta^3}}}\sinh\left(\frac{\beta t}{2}\sqrt{1+\frac{4D\lambda^2}{\beta^3}}\right)\right)\nonumber\\&
+\mathcal{O}\left(\left(\frac{D\lambda^2}{\beta^3}\right)^2\right) \,.
\elabel{DysonSolution}
 \end{align}
The basic 
 reproduction number $R_0$ that 
 results in asymptotically constant expected population size, $\infty>\lim_{t\to\infty} \mean{N(t)}>0$,
 according to \eref{DysonSolution}, is the one that makes the exponent vanish for all $t$, namely $s(1-R_0)=\beta/2(\sqrt{1+4D\lambda^2/\beta^3}-1)$, or
 \begin{subequations}
 \begin{align}\label{eq_critical_point_OU_noise}
    R_0&=1+\frac{\beta}{2s}\left({1-\sqrt{1+\frac{4D\lambda^2}{\beta^3}}}\right)+\mathcal{O}\left(\left(\frac{D\lambda^2}{\beta^3}\right)^2\right)\\
    &=1-\frac{D\lambda^2}{\beta^2 s} +\mathcal{O}\left(\left(\frac{D\lambda^2}{\beta^3}\right)^2\right).
\end{align}
 \end{subequations}
Remarkably, in any non-trivial setup, this critical $R_0$ is less than unity. The average number of secondary infections of an isolated individual not subject to noise needed to sustain an outbreak, is thus less than unity. The explanation for this counter-intuitive result is similar to the reason why \Eref{R0_Dawson_expanded} is based on the wrong assumptions: Because the noise is correlated, in general population sizes experiencing low extinction rates are larger than those experiencing large extinction rates. 
The noise correlator of the the Ornstein-Uhlenbeck process 
\eqref{epsilon_OU} is \cite{vanKampen:1992}
\begin{equation}
\elabel{OU-correlation}
\mean{y(t)y(t')}=\frac{D}{\beta} \exp{-\beta |t-t'|} \,,
\end{equation} 
whose characteristic time $\beta^{-1}$  is a measure of the persistence of active fluctuations.
Although the noise has vanishing mean, its effect is biased towards larger population sizes. 
In addition, Eq.~\eref{eq_R_0_OU} does not incorporate the change in frequency with which events take place overall --- at times of high extinction rates, more events take place than at times of low extinction rates.

Our field theoretic result Eq.~\eqref{eq_critical_point_OU_noise} provides a systematic expansion of the critical $R_0$ in orders of $\lambda$ and is subtly different from the \emph{ad hoc} result \eref{R0_Dawson_expanded}, as the denominator of the leading order correction in \eref{R0_Dawson_expanded} is $\beta s^2$ rather than $\beta^2 s$ in Eq.~\eqref{eq_critical_point_OU_noise}.

\begin{figure}
    \centering
    \includegraphics[width=0.5\textwidth]{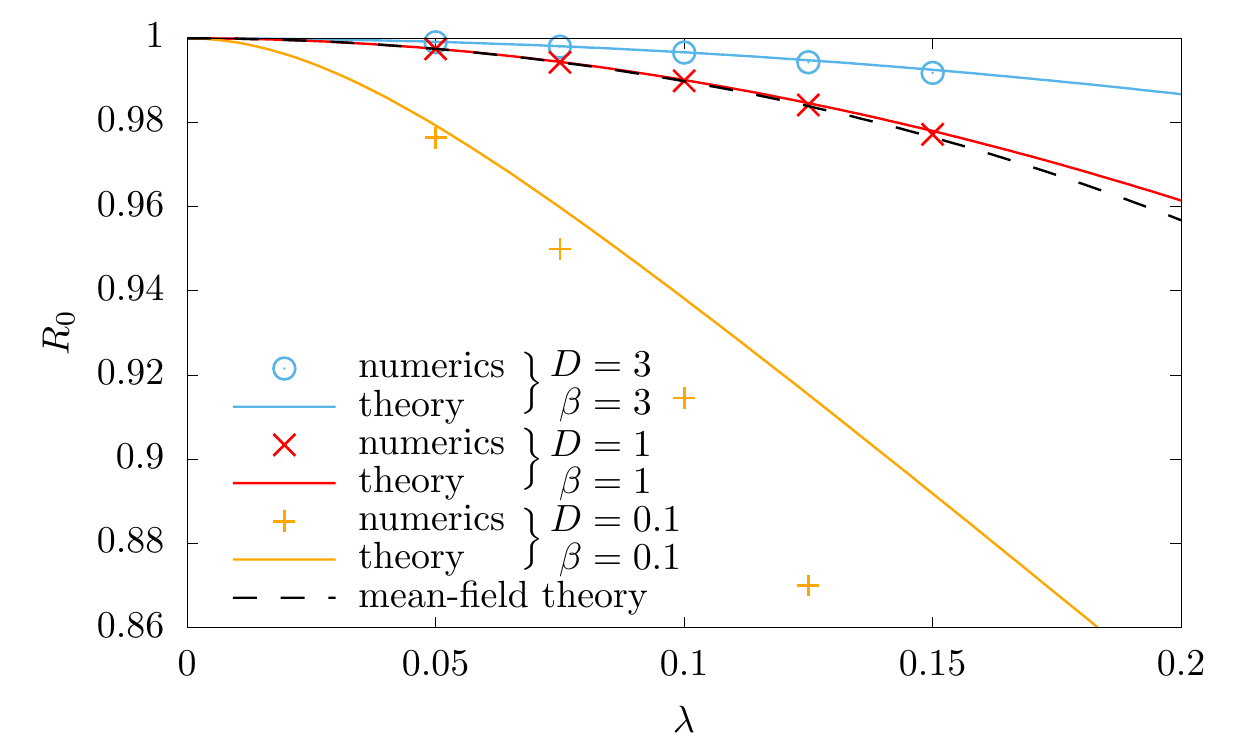}
    \caption{
    Phase diagram of branching process coupled to an 
    Ornstein-Uhlenbeck process for $s=1$ and different values of $D$ and $\beta$.
    We estimated numerically the critical $R_0$  for
    some values of $\lambda$ (symbols), see 
    Fig.~\ref{sample_OU_trajectories}.
    Critical values separate
    subcritical (below) and 
    supercritical regimes (above). 
    Solid lines (labelled as "theory") indicate the 
    critical $R_0$ for a given $\lambda$
    as approximated by  Eq.~\eqref{eq_critical_point_OU_noise}.
    These theoretical curves are first-order approximations in orders of $D\lambda^2/\beta^3$, which explains
    the deviation of the numerical estimates from the theory. We expect larger deviations between true 
    values and our approximation \eqref{eq_critical_point_OU_noise}
    for larger ratios of $D/\beta^3$.
    The mean-field approximation in \eref{R0_Dawson_expanded} (dashed line) is common to the three sets of values since $D/\beta$ is the same in all three cases.
    }
    \label{fig:critical_point_numerics}
\end{figure}

To test Eq.~\eqref{eq_critical_point_OU_noise} numerically, we have performed Monte-Carlo simulations to estimate the critical $R_0$ as shown in Fig.~\ref{fig:critical_point_numerics}. Given the other parameter values, a fairly small range of $\lambda$ is available, as otherwise $\epsilon(t)$ might stray in to negative territory. The perturbative result Eq.~\eqref{eq_critical_point_OU_noise} is in excellent agreement with the numerics. Fig.~\ref{sample_OU_trajectories} shows some examples of subcritical, near critical and supercritical trajectories of $\mean{\Num(t)}$.

\section{Extinction Rate coupled to Telegraphic noise}
\label{Sec:TelegraphicNoise}
As a second example of how noise can shift the critical point in unexpected ways, we consider a branching process in which the extinction rate switches spontaneously between two values. We call this random switching  telegraphic noise \cite{HorsthemkeLefever:1989} and write,
\begin{equation}\elabel{telegraphic_noise_intro}
    \epsilon(t)= sp_0 + T(t)\,,
\end{equation}
where the binary random variable $T$ switches between
the two values $\eon>0$ and $\eoff=0$. An example trajectory is shown in Fig.~\ref{fig:traj_BP_T}. Analogously to Eq.~\eqref{eq_time_dependent_R_OU}, we can immediately deduce the time-dependent expected number of secondary cases\begin{align}\label{eq_time_dependent_R_Tel}
\mean{K(t)}=\frac{s}{s+T(t)}R_0\,,
\end{align}
with $R_0$ as defined by the distribution $p_k$, Eq.~\eqref{def_R0}.
The waiting times between switching events are exponentially distributed with rates 
$\moff$ to go from $\eon$ to $\eoff$, and
$\mon$ to go from state $\eoff$ to $\eon$,
\begin{align}
\elabel{tele_reactions}
\eon
\xrightarrow[]{\moff}  
\eoff \,,&&
\eoff
\xrightarrow[]{\mon} 
\eon \,.
\end{align}
The switching rates $\mon$ and $\moff$ induce temporal correlations in the noise $T(t)$ in the same way as the active fluctuations in
the motion of run-and-tumble particles \cite{DharETAL:2019,Garcia-MillanPruessner:2021}, \Eref{telegraphic-noise-correlation}.


\begin{figure*}
\includegraphics[width=\textwidth]{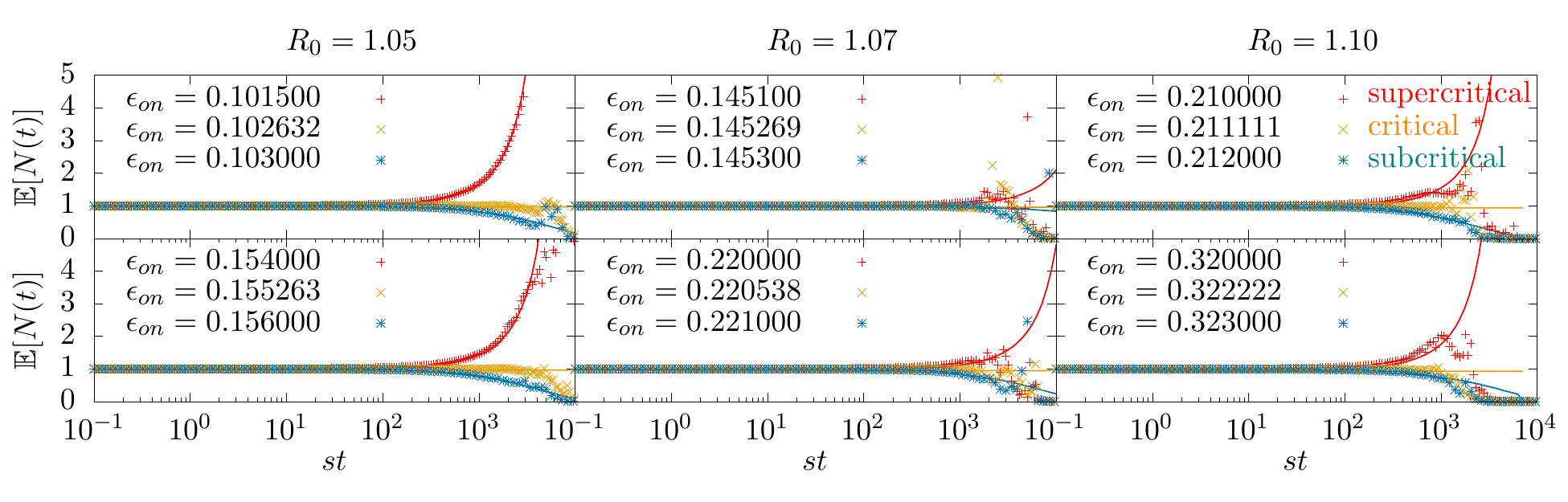}
\caption{ \label{sample_Tel_trajectories}
Expected number of infected individuals $\mean{N(t)}$ of a branching process
driven by a dichotomous Telegraphic  noise
estimated from numerical simulations (symbols) based on $10^7$ iterations.
The parameters of the noise are as follows: equal switching rates $\mon = \moff = 1$ 
(top row), and unequal switching rates $\mon = 1$ and $\moff = 2$ (bottom row).
For each set of noise parameters, we can choose different values
of $\eon$ such that the population has 
supercritical (red symbols, black lines),
critical (orange symbols, grey lines),
and subcritical dynamics (blue symbols, pink lines).
In all instances, $\eoff=0$.
Our numerical results are in agreement with \Eref{tele_ave_N}.}
\end{figure*}


The master equation describing the telegraphic noise only is
\begin{subequations}\label{eq_master_telegraphic}\begin{align}
    {\partial_t}P(\eon;t)=&\mon P(\eoff;t)-\moff P(\eon;t),\\
    {\partial_t}P(\eoff;t)=&\moff P(\eon;t)-\mon P(\eoff;t)
\end{align}\end{subequations}
where $P(T;t)$ is the probability distribution of $T$  at time $t$. 
From Eq.~\eqref{eq_master_telegraphic} we derive the expected value of $T$ at time $t>0$:\begin{align}
\mean{T(t)}=\mean{T(0)}e^{-(\mon+\moff)t}+\frac{\eon\mon}{\mon+\moff}.  
\end{align}
In this set-up, the expected number of secondary cases $\mathbb{E}[K]$, Eq.~\eqref{eq_time_dependent_R_Tel}, switches between two values. If one of them predicts supercritical behaviour and the other one predicts subcritical behaviour, we cannot immediately deduce the criticality of the population that randomly switches between them.

One way of determining the critical point is to demand that the rate with which offspring are \emph{produced} in branching events equals that with which they go extinct.
In each branching into $k$ particles, $k-1$ offspring are produced, so that the production rate of particles is
\begin{equation}\label{eq:telegraphic_production}
    s \left( \sum_{k=2}^\infty (k-1) p_k\right)= s (R_0+p_0-1) 
\end{equation}
using Eq.~\eqref{def_R0}. This \emph{production} rate is unaffected by the state of the system. The extinction, on the other hand, depends on the state.
As the transitioning times are exponentially distributed, the system spends on average $1/\mon$ amount of time in the $T=\eoff=0$ state and $1/\moff$ amount of time in the $T=\eon$ state.
This implies that at an arbitrary point in time, the system is in state $T=\eoff=0$ with probability $\moff/(\mon+\moff)$ and 
in state $T=\eon$
with probability $\mon/(\mon+\moff)$.
The effective extinction rate is therefore
\begin{equation}
    sp_0 \frac{\moff}{\mon+\moff}
    + (sp_0+\eon) \frac{\mon}{\mon+\moff}
    = sp_0 + \frac{\eon\mon}{\mon+\moff}\ .
\end{equation}
Equating this with Eq.~\eqref{eq:telegraphic_production} produces the criterion for the critical point,
\begin{align}\label{eq_MF_tele}
    \frac{\eon}{s}=(R_0-1)\left(1+\frac{\moff}{\mon}\right) \ .
\end{align}

However, Eq.~\eqref{eq_MF_tele} does not correctly predict the critical point as generally a larger population is affected by small extinction rates than by large extinction rates, because $\mon$ and $\moff$ are finite, so that the system lingers in either state. The mean field theory is expected to describe only the case of $\mon,\moff\gg s$ correctly, when the telegraphic noise changes so quickly that population size and state of the noise become uncorrelated. In its steady state, the telegraphic noise has a Pearson correlation coefficient $\rho$ of\begin{align}
    \rho_{T(t)T(t')}=&\,\frac{\mean{T(t)T(t')}-\mean{T(t)}\mean{T(t')}}{\sqrt{\mathbb{V}[T(t)]\mathbb{V}[T(t')]}}\nonumber\\
    =&\,e^{-(\mon+\moff)|t-t'|},\elabel{telegraphic-noise-correlation}
\end{align}
which indicates that correlations become irrelevant as $\mon$ and $\moff$ become large compared to $s$, the other event rate of the system. 
The derivation of \Eref{telegraphic-noise-correlation} is presented in Appendix~\ref{app-telegraphic-noise-correlation}.

As in Sec.~\ref{Sec:LinearOU}, the expectation of the number of offspring averaged over all branching events is not a simple average of Eq.~\eqref{eq_time_dependent_R_Tel}, as it lacks a weighting by population size.



\begin{figure}
    \centering
    \includegraphics[width=\columnwidth]{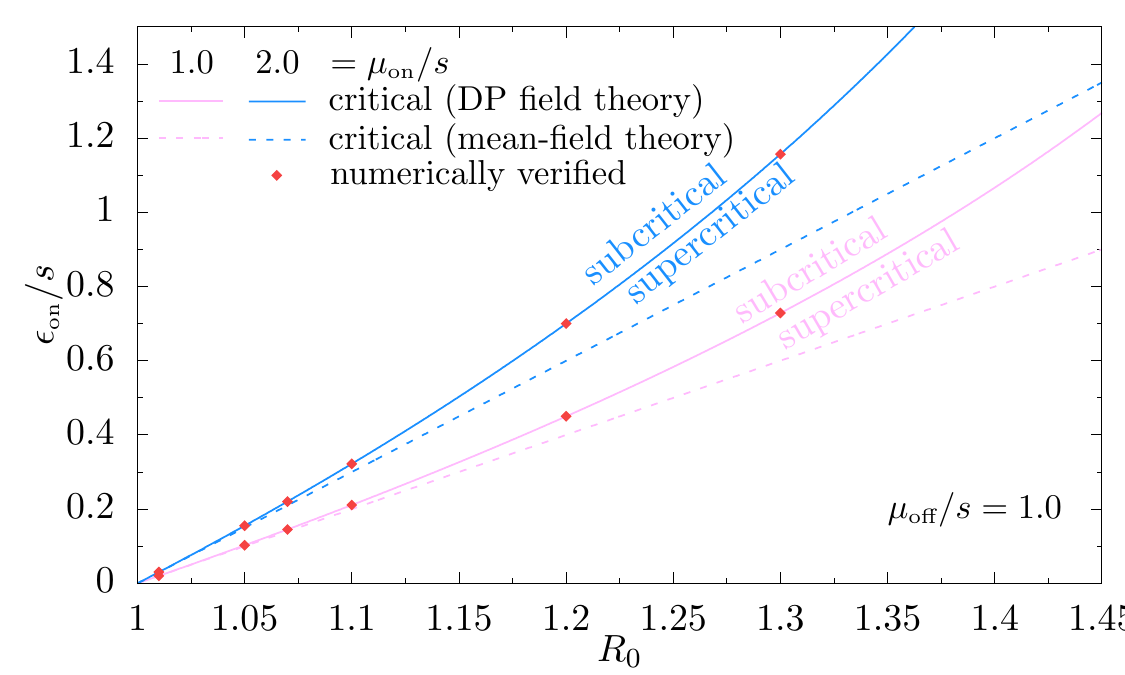}
    \caption{Two phase diagrams of branching process coupled to telegraphic noise (blue and pink). The critical $R_0$-$\epsilon_\textit{on}/s$ line separating the subcritical (above left) from the supercritical regime (below right). Solid line: exact result, Eq.~(\ref{eq_tele_crit}); dashed line: mean-field result, Eq.~\eqref{eq_MF_tele}. The red dots indicate where the critical line was verified by Monte-Carlo simulations, see Fig.~\ref{sample_Tel_trajectories}.}
    \label{fig:telegraphic-regimes}
\end{figure}

In order to capture whether outbreaks are supercritical or not, we inspect again the expected number of infected individuals over time $\mean{N(t)}$. For the calculation of $\mean{N(t)}$, we use a Doi-Peliti field theory, derived in Appendix~\ref{Sec:appendix-telegraphic-noise}, with initial condition $N(0)=1$ and $T(0)=\eon$. 
We determine in which parameter region $\mean{N(t)}$ grows over time (supercritical phase) and which it decreases (subcritical phase). The boundary between the two regions defines the critical hypersurface
\begin{equation}\label{eq_tele_crit}
\frac{\eon}{s}=(R_0-1)\left(1+\frac{\mu_\textit{off}}{s(1-R_0)+\mu_\textit{on}}\right)\,,
\end{equation}
which is shown in Fig.~\ref{fig:telegraphic-regimes}.

The direct comparison between the mean-field approach, Eq.~\eqref{eq_MF_tele}, and the exact result, Eq.~\eqref{eq_tele_crit}, in Fig.~\ref{fig:telegraphic-regimes} shows that the mean-field approximation predicts subcritical behaviour in regions where the dynamics are actually supercritical, \ie~in the regions between solid and dashed lines. As expected, Eqs.~\eqref{eq_MF_tele} and \eqref{eq_tele_crit} coincide when $\mon,\moff\gg s$, in which case they reduce to $\eon/s=(R_0-1)(1+\moff/\mon)$.

We verified the shifted critical line using Monte-Carlo simulations, shown in Fig.~\ref{sample_Tel_trajectories}. 

\section{Coupled Branching Processes}
\label{Sec:Coupled-Birth-Death-Process}

While in the previous sections the branching process was coupled to different noises via a dynamic change in the extinction rate, here we study a type of noise
introduced by the interaction between different populations. In the context of infectious diseases, we can think of population subgroups with different susceptibility, 
perhaps as a matter of lifestyle, behaviour or underlying health condition.
As before, we leave various interpretations and applications to the reader and focus on analyzing the dynamics of an example process. 

We consider a branching process that is coupled to another branching process. 
Individuals from two populations $A$ and $B$ with 
branching probabilities $p_{kA}$ and $p_{kB}$, Eq.~\eref{def_p_k},
respectively, change from one population to the other with
 transmutation rates $\muA$ (from $A$ to $B$) and $\muB$ (from $B$ to $A$),
\begin{align}
\elabel{AB_interaction}
A
\xrightarrow[]{\muA}  \,
B \,, &&
B
\xrightarrow[]{\muB}  \,
A \,.
\end{align}
An example trajectory is shown in Fig.~\ref{fig:traj_coupledBP}. The master equation of this process involves the joint 
probability $P(N_A,N_B;t)$, where $N_A$ and $N_B$ are the number
of individuals of populations $A$ and $B$ respectively at time
$t$. The master equation is made of three blocks describing
each subprocess: two independent branching processes for
sub-populations $A$ and $B$, modelled in 
\eqref{eq_master_branching}
and a coupling term that models the interaction 
\eref{AB_interaction} between the two populations,
\begin{align}
    {\partial_t}P(N_A,N_B;t)
    =  &   {\partial_t}P_A(N_A,N_B;t)
    + {\partial_t}P_B(N_A,N_B;t) \nonumber\\
    & + {\partial_t}P_{AB}(N_A,N_B;t) \,.
    \elabel{eq_master_coupled}
\end{align}
 Here, 
 the term ${\partial_t}P_A(N_A,N_B;t)$ is given by \eqref{eq_master_branching} 
 replacing $N$ by $N_A$, $p_k$ by $p_{kA}$ and $P(N;t)$ by $P(N_A,N_B;t)$;
 the term ${\partial_t}P_B(N_A,N_B;t)$ is given by \eqref{eq_master_branching} 
 replacing $N$ by $N_B$, $p_k$ by $p_{kB}$ and $P(N;t)$ by $P(N_A,N_B;t)$;
 and the term ${\partial_t}P_{AB}(N_A,N_B;t)$ captures the transmutation of indviduals in \eref{AB_interaction},
\begin{align}
    &{\partial_t}P_{AB}(N_A,N_B;t) \notag\\
    & = \muA \bigl((N_A+1)P(N_A+1,N_B-1;t)
   -N_A P(N_A,N_B;t)\bigr)\notag\\
    &\,+\muB \bigl((N_B+1)P(N_A-1,N_B+1;t)
    -N_B P(N_A,N_B;t)\bigr)\,.
\end{align}
This coupling term describes how an individual of population $A$ joins population $B$ with rate $\muA$ and how individuals from $B$ convert to $A$ with rate $\muB$.


\begin{figure*}
\includegraphics[width=\textwidth]{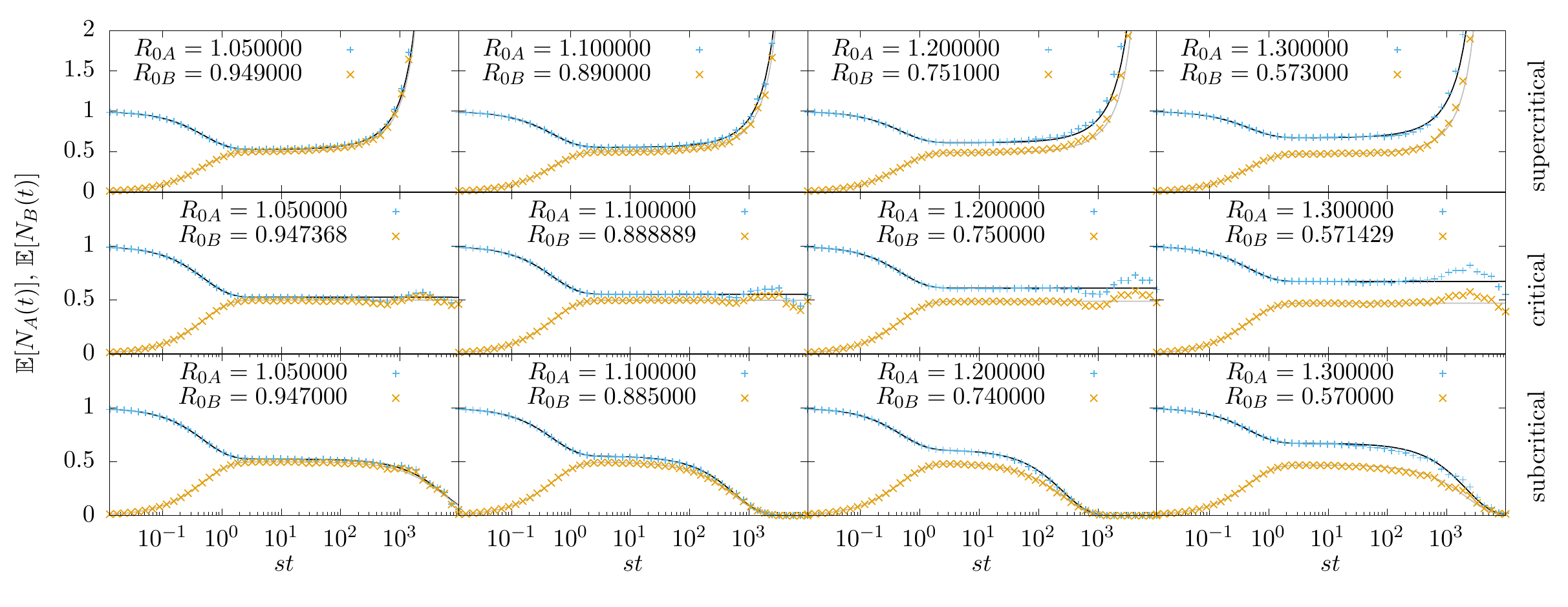}
\caption{\label{sample_BPBP_trajectories1}
Expected number of individuals in two coupled
branching processes
$\mean{N_A(t)}$ (blue symbols, black lines) and  $\mean{N_B(t)}$ (orange symbols, grey lines).  
The transmutation rates are $\muA=\muB=1$ (\cf Fig.~\ref{sample_BPBP_trajectories2} for $\muA=1$ and $\muB=2$).
The basic reproduction number $R_{0A}$
is above $1$ and increases from left to right
panels. The basic reproduction number 
$R_{0B}$ is adjusted in each panel to illustrate
supercritical processes (top row),
critical processes (middle row),
and subcritical processes (bottom row).
The supercritical cases shown in the top row are incorrectly predicted to be subcritical by the mean-field theory.
Numerical estimates (symbols), based on 
$2\cdot 10^5$ trajectories, are in good agreement with
exact predictions (lines) in \Eref{exp_N_BP_BP}.
}
\end{figure*}

\begin{figure*}
\includegraphics[width=\textwidth]{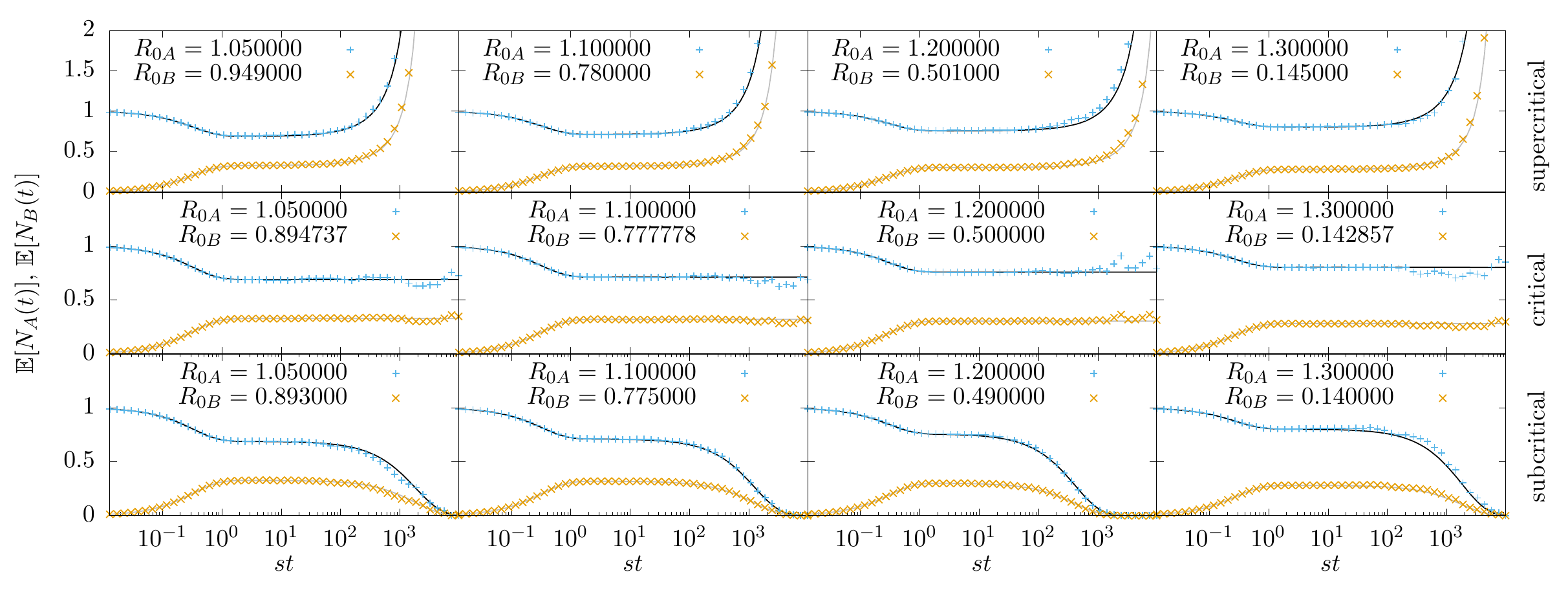}
\caption{\label{sample_BPBP_trajectories2}
Expected number of individuals in two coupled
branching processes
$\mean{N_A(t)}$ (blue symbols, black lines) and  $\mean{N_B(t)}$ (orange symbols, grey lines).  
The transmutation rates are $\muA=1$ and $\muB=2$ (\cf Fig.~\ref{sample_BPBP_trajectories1} for $\muA=\muB=1$).
The basic reproduction number $R_{0A}$
is above $1$ and increases from left to right
panels. The basic reproduction number 
$R_{0B}$ is adjusted in each panel to illustrate
supercritical processes (top row),
critical processes (middle row),
and subcritical processes (bottom row).
The supercritical cases shown in the top row are incorrectly predicted to be subcritical by the mean-field theory.
Numerical estimates (symbols), based on 
$2\cdot 10^5$ trajectories, are in good agreement with
exact predictions (lines) in \Eref{exp_N_BP_BP}. }
\end{figure*}


To derive the dynamics of one of the two populations, say $A$,
we marginalise the joint probability $P(N_A,N_B;t)$ by summing
over $M$, which gives the probability that population $A$ has
$N_A$ individuals at time $t$,
\begin{equation}
    P(N_A;t)=\sum_{N_B\geq0}P(N_A,N_B;t).
\end{equation}
By summing over $N_B$ in \eref{eq_master_coupled} we obtain 
\begin{align}\elabel{P_A_marginalised}
    {\partial_t}
    P(N_A;t)
    =& s\sum\limits_{k=0}^\infty p_{kA}(N_A-k+1)P(N_A-k+1;t)\notag\\
    &-s N_A P(N_A;t)\notag\\
    &+\muA\bigl((N_A+1)P(N_A+1;t)
    -N_A P(N_A;t)\bigr)\notag\\
    &+\muB\bigl(\mean{N_B(t)|N_A(t)-1}P(N_A-1;t)\notag\\
    &-\mean{N_B(t)|N_A(t)}P(N_A;t)\bigr) \,,
\end{align}
which shows that, from the perspective of sub-population $A$,
its dynamics can be cast into a branching process \eqref{eq_master_branching} with slightly adjusted extinction rate and an additional influx, akin to spontaneous creation. The first two terms on the right hand side of Eq.~\eref{P_A_marginalised} are indeed identical to the branching process in Eq.~\eqref{eq_master_branching}, the term parameterised by $\muA$ corresponds to a spontaneous extinction and the last term, parameterised by $\muB$, is reminiscent of a spontaneous creation. However, the rates of the gain and loss terms of this creation differ and depend on $\mean{N_B(t)|N_A(t)}$, which is a deterministic function of the stochastically varying size $N_A$ of sub-population $A$. It is the only term that links the dynamics of the two sub-populations. In particular, it encapsulates the conservation of individuals by transmutation.
The branching dynamics of sub-population $B$ disappears from the dynamics of sub-population $A$ otherwise.

If the branching processes of both populations are supercritical, we expect the coupled populations to remain supercritical, irrespective of the transmutation, as it conserves the total population size and cannot reduce it. Similarly if both processes are subcritical, as the transmutation cannot increase the population size either. However, the overall dynamics is not 
straightforward if the two populations lie in different 
criticality regimes.
Without loss of generality, we assume in the following that 
basic reproduction numbers are $R_{0A}>1$ and $R_{0B}<1$, both defined by Eq.~\eqref{def_R0} with $p_k$ replaced $p_{kA}$ and $p_{kB}$ respectively. Is the joint population of $A$ and $B$ super- or subcritical?

\begin{figure}
    \centering
    \includegraphics[width=\columnwidth]{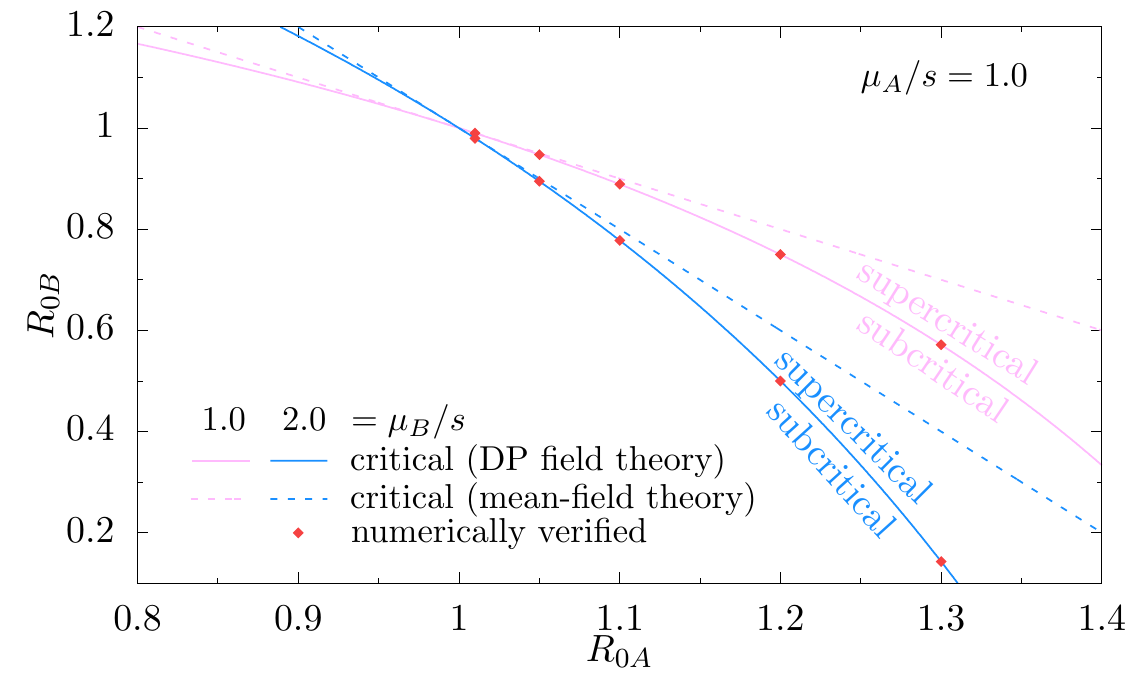}
    \caption{Two phase diagrams of coupled branching processes (blue and pink). The critical $R_{0A}$-$R_{0B}$ line between supercritical (top right) and subcritical (bottom left) regimes for $\muA/s=\muB/s=1$ (pink) and $\muA/s=1$, $\muB/s=2$ (blue). A mean-field approach would predict the critical line to equal the dashed line. The red dots indicate where the critical line was verified by Monte-Carlo simulations, see Figs.~\ref{sample_BPBP_trajectories1} and~\ref{sample_BPBP_trajectories2}.}
    \label{fig:coupled-birth-death-regimes}
\end{figure}

A naive approach would be to consider how much time an individual spends as part of population $A$ before joining population $B$ and vice versa. As the waiting time between transmutations is exponentially distributed with parameters $\muA$ and $\muB$ respectively, an individual spends on average $1/\muA$ time in population $A$ and $1/\muB$ time in population $B$. Thus, a weighted average of the two $R_0$ values is given by
\begin{equation}
    \frac{R_{0A}/\muA + R_{0B}/\muB}{1/\muA+1/\muB} = 
    \frac{R_{0A}\muB + R_{0B}\muA}{\muA+\muB}
\end{equation}
and demanding that this weighted average is unity at the critical point determines the
critical hypersurface as
\begin{align}\label{eq:MF-result-coupled-branchers}
R_{0B}=1+\frac{(1-R_{0A})\muB}{\muA},
\end{align}
which is shown in Fig.~\ref{fig:coupled-birth-death-regimes} as dashed lines. This \emph{ad hoc} approximation ignores some important details of the interaction between the two sub-populations:
Firstly, many more individuals are initiated in sub-population $A$, a bias not accounted for by the time-averaging taken above. Secondly, whenever a individual resides in $A$, many more branching events, namely those of its many offspring, will be characterised by $R_{0A}$. As in 
Sec.~\ref{Sec:TelegraphicNoise}, only in the limit of large $\muA,\muB$ with constant $\muA/\muB$, can we expect Eq.~\eqref{eq:MF-result-coupled-branchers} to be correct.

In order to find the critical point where the average total population size starts displaying exponential growth, we use a Doi-Peliti field theory, Appendix~\ref{app-coupled-branchers}. 
Given transmutation rates $\muA$ and $\muB$, the border between the subcritical and the supercritical regime is 
the critical line  defined by
\begin{align}\label{eq:critical_line_coupled_branchers}
    R_{0B}=&1+\frac{(1-R_{0A})\muB}{s(1-R_{0A})+\muA}\,,
\end{align}
shown in Fig.~\ref{fig:coupled-birth-death-regimes} with
solid lines. Indeed, the mean-field result Eq.~\eqref{eq:MF-result-coupled-branchers} is recovered in the limit of $\muA,\muB\to\infty$ at constant $\muA/\muB$, when $R_{0B}=1+(1-R_{0A})\muB/\muA$.

As in the previous sections (Figs.~\ref{fig:critical_point_numerics} and \ref{fig:telegraphic-regimes}),
Fig.~\ref{fig:coupled-birth-death-regimes} 
shows that
the mean-field approach, Eq.~\eqref{eq:MF-result-coupled-branchers}, may estimate subcritical behaviour where an exact calculation reveals supercritical dynamics, \ie~in the regions between solid and dashed lines.
The shift of the critical line Eq.~\eqref{eq:critical_line_coupled_branchers} is confirmed in Monte-Carlo simulations, shown in Figs.~\ref{sample_BPBP_trajectories1} and~\ref{sample_BPBP_trajectories2}.
\section{Discussion and Conclusion}
\label{Sec:Conclusion}
Branching processes are used to model a variety of avalanche-like dynamics ranging from neuronal activity to infectious diseases. One of the key points of interest is the prediction of the criticality of the dynamics, \ie~whether activity might diverge and continue forever or whether it will die out eventually. Here, we show in several examples that global noise in the branching  dynamics  induce correlations in the entire population which influence the growth or decline of the population over time. 

The three examples are 
i) Sec.~\ref{Sec:LinearOU}, a branching process in which the subprocess of extinction follows an Ornstein-Uhlenbeck process, 
ii) Sec.~\ref{Sec:TelegraphicNoise}, a branching process in which the subprocess of extinction displays telegraphic noise, \ie~its extinction rate randomly switches between two values and 
iii) Sec.~\ref{Sec:Coupled-Birth-Death-Process}, two coupled branching processes where individuals can convert from one branching process and its parameters to another branching process with a different set of parameters. 

In each case, we saw that mean-field arguments fail, even when they seem to capture much of the dynamics. In each case, correlations and fluctuations are important to be captured correctly, something that coarse-grained descriptions, such as SIR models, routinely ignore. Instead, we made use of more sophisticated, field-theoretical techniques to characterise the branching processes either perturbatively, as in the case of Ornstein-Uhlenbeck noise (Secs.~\ref{Sec:LinearOU} and \ref{appOU}) or exactly (Secs.~\ref{Sec:TelegraphicNoise}, \ref{Sec:Coupled-Birth-Death-Process}, \ref{Sec:appendix-telegraphic-noise} and \ref{app-coupled-branchers}). This approach allowed us to determine effective critical values of $R_0$ that, surprisingly, are less than unity.


\subsection{Failure of Mean-Field Theories}
In all three examples, the effect of the noise is that the critical point dividing divergence from decline is shifted in unexpected ways. Mean-field approaches predict subcritical behaviour in some parameter regions where in fact supercritical dynamics occur.
The reason for the failure of the mean-field theory is that it is based on averages that do not take the population size into account. When the extinction rate fluctuates, typically large populations are exposed to low extinction, and typically small populations are exposed to high extinction rates, unless the change of the extinction is fast compared to the process. 
And yet, at the critical point, as gains and losses are balanced, we expect the typical population sizes to be identical in the low and the high extinction rate regimes. In other words, according to this argument, the mean-field theory may fail to characterise the average number of offspring produced, but it should still predict the critical point correctly. However, critical or not, small, symmetric fluctuations in the time spent in a state of either high or low extinction rate have a disproportionate effect on the population size: Spending additional time $\Delta t$ in a state of low extinction creates many more offspring, given the initially large population size, than go extinct by spending the same additional time $\Delta t$ in the state of high extinction, or that are failed to be created when the time spent in the low extinction rate is reduced by $\Delta t$. 

A similar bias is visible in the expectation of the exponential 
$\mathbb{E}[ \exp{z_t}]=1+\mathbb{E}[z_t^2]+\ldots>1$ of a symmetric random walk $z_t$ with $\mathbb{E}[z_t]=0$. Somewhat \emph{ad hoc}, the population size of the branching process with Ornstein-Uhlenbeck noise, Sec.~\ref{Sec:LinearOU} and Sec.~\ref{appOU}, may in fact be approximated by assuming that it is the exponential of the time integral of the instantaneous effective mass $r$ \cite{Garcia-Millan2018}, and correspondingly averaged,
\begin{subequations}
\begin{align}
    &\mathbb{E}[ N(t) ] \approx \mathbb{E}\left[ \Exp{-\int_0^t (r + \lambda y(t')) \plaind t'} \right]\\
    &= \exp{-rt} \Big\{1+\frac{\lambda ^2}{2}\mathbb{E}\left[ \left( \int_0^t y(t') \plaind t' \right)^2\right]
    + \ldots \Big\}
\end{align}
\end{subequations}
as the total, effective, instantaneous mass is $r + \lambda y_{t}$ with $r=s(1-R_0)$,  \Eref{all_def_r}, and $\mathbb{E}[ y(t) ]=0$. Using $\mathbb{E}[ y(t) y(t') ]=(D/\beta) \exp{-\beta |t-t'|}$ of the Ornstein-Uhlenbeck process \cite{vanKampen:1992}, this integral produces indeed the correct first order correction, \Eref{one_loop_correction},
\begin{align}
    &\mathbb{E}[ N(t) ] 
    \approx  \exp{-rt} \Big\{1+\frac{D \lambda ^2}{\beta^3} 
    (\exp{-\beta t}-1 + \beta t) 
    + \ldots \Big\} \ ,
\end{align}
and, apparently, all higher order corrections of the population size, Eq.~\eqref{eq:linearOU-OneLoopCorrections}. 
Given the \emph{ad hoc} nature of this expression, we cannot confirm its validity to all orders, nor do we know whether correlation functions are correctly captured \cite[\cf Eqs.~(6) and (32) in][]{Pausch2020}.


\subsection{Implications for infectious disease modelling}
In this paper we consider a branching process with an external noise
as a basic model of infectious disease spreading.
This model of unmitigated epidemic assumes an infinite population of 
susceptible individuals and hence it does not account for elements
such as immunity or saturation. However, our results show
an important aspect in disease spreading that, 
to our knowledge, has not been accounted for
before \cite{Andreasen:2011}: \emph{fluctuations present in the transmission of the infection
can shift the critical value of $R_0$ below $1$}.
Therefore, we may observe an epidemic outbreak despite $R_0$ being
smaller than $1$. 
This needs to be taken into account when
designing interventions aimed at containing
an epidemic outbreak.

Moreover, we expose the failure of mean-field theories, which are widely used in epidemic modelling, to predict 
the critical point, see discussion above. 
The main reason for this failure is that
mean-field theories and deterministic models, such as the classic
SIR model, are not designed for capturing fluctuations and noise.
These may provide useful approximations in the mist of an ongoing
epidemic crisis \cite{GogHollingsworth:2021,GogETAL:2021,KucharskiETAL:2020,AbbottETAL:2020,RiccardoETAL:2020} but lack the capacity
to account for randomness.
Therefore, incorporating the stochastic nature of
disease spreading in epidemiologic models will provide a better
understanding and prediction of the evolution of an outbreak \cite{ArielLouzoun:2021}.
We leave for future work the study of the effect of 
immunisation and the study of interventions that are able to contain the outbreak
as well as the  calculations of observables such as the peak prevalence (maximum value
of infected individuals over time) and the final size of the outbreak
(proportion of population infected at any time during the epidemic).



\subsection{Conclusion and Outlook}



In practice, the failure of the mean-field theory means that a faithfully defined critical $R_0$ value which is obtained by ignoring fluctuations and correlations is unreliable. Even basic, handwaving arguments fail --- we had initially expected the noise to push the critical value of $R_0$ above unity, because additional fluctuations might terminate a branching process by wiping out the last individual of a small, highly volatile population of an otherwise near-critical branching process. Instead, as long as the noise has a finite correlation time, so that the system \emph{lingers} in a state of higher or lower extinction rate, there is an intrinsic bias towards larger populations.
The most striking consequence of this bias is the critical basic reproduction number $R_0$, in our naive definition Eq.~\eqref{def_R0}, becoming less than unity. 
There is no unique definition of $R_0$, yet in the case of unbiased Ornstein-Uhlenbeck noise, Sec.~\ref{Sec:LinearOU}, we cannot think of a redefinition of $R_0$ that renders its critical value unity.

Future research will focus on finding a more suitable observable or set of observables which constitute a sufficient predictor for the criticality of the noisy branching dynamics.

\section*{Acknowledgements}
We thank Andy Thomas for his brilliant technical support, Nanxin Wei and Guillaume Salbreux for fruitful discussions. J.P.~was supported by an EPSRC Doctoral Prize Fellowship.
R.G.M.'s~work was funded by the European Research Council under the EU’s Horizon 2020 Programme, grant number 740269.

\section*{Author's Contributions}
All authors were involved in all aspects of the conceptualization and investigation as well as the writing and editing of the manuscript. JP acquired partial funding for this project through an EPSRC Doctoral Prize Fellowship. Data visualization and other graphics were created by RGM and JP. RGM and GP acquired computing resources through Imperial College London. 

\appendix
\newcommand{\tildechi}{\widetilde\chi}
\newcommand{\tildepsi}{\widetilde\psi}

\section{Branching process field theory}
\label{app_branching_field_theory}
We recall from \cite{Garcia-Millan2018} the Doi-Peliti field theory for the continuous-time branching process with arbitrary time-independent offspring distribution, where $K\in\{0,1,\dots\}$ is the number of offspring  produced at a branching event with probability $P(K=k)=p_k$, \Eref{def_p_k}. The waiting times between two branching events is exponentially distributed with rate $s$. The action of this field theory
is
\begin{subequations}
\elabel{all_def_r}
\begin{align}
\label{eq:birth-death-action}
    \mA_{\text{BP}}[\phi,\widetilde\phi]=&\int\dint{t}\left\{-
    \widetilde\phi\left(\partial_t+r\right)\phi+\sum\limits_{j=2}^\infty q_j\widetilde\phi^j\phi\right\},\\
    \text{with }r=&s\left(1-R_0\right)\text{ and }q_j=s\sum\limits_{k=0}^\infty {k\choose j}p_k,\elabel{all_def_r_partii}
\end{align}
\end{subequations}
where the fields $\phi$ and $\widetilde\phi$ are functions of time $t$ and the binomial coefficient is zero if $j>k$. 
We assume that the system is initialised with one 
individual at time $t_0=0$, $N(t_0)=1$, throughout.
To calculate an observable ${\mO}$ in the dynamics of the 
branching process, we 
perform the path integral 
\begin{equation}
\ave{\mO}_{\text{BP}}=\int\mD\phi\mD\tildephi\,\, \mO \,\,
\exp{\mA_{\text{BP}}[\phi,\tildephi]}\,,
\end{equation}
which generally involves the bilinear part in \eref{birth-death-action}
and the nonlinear couplings
$q_j$.
However, the observables that we are concerned with in this
paper, do not involve the couplings $q_j$, so we do not consider
them beyond this point. The Gaussian model of the branching process
follows from the bilinear part of \eref{birth-death-action}, $\mA_0[\phi,\widetilde\phi]=-\int\dint{t}\widetilde\phi\left(\partial_t+r\right)\phi$, which gives
 the bare propagator from the Gaussian model
\cite{Garcia-Millan2018},
\begin{equation}
    \tikz[baseline=-2.5pt,scale=0.75]{
\draw[Aactivity] (0:-0.8) -- (0:0.8);
}\corresponding
\ave{\phi(\omega)\tildephi(\omega')}_0 =
\frac{\deltabar(\omega+\omega')}{-\imag\omega+r} \,,
\elabel{BPbareprop}
\end{equation}
where  $\deltabar(\omega+\omega') = 2\pi \delta(\omega+\omega')$ and\begin{align}
    \ave{\bullet}_0=\int\mathcal{D}\phi\mathcal{D}\tildephi\dots\bullet e^{\mathcal{A}_0[\phi,\tildephi,\dots]}.
\end{align}
In \eref{BPbareprop}, the frequencies 
$\omega$ and $\omega'$ are reciprocal to times $t$ and $t_0$ under the Fourier transform
convention,
\begin{align}
    \phi(\omega) = \int\dint{t} \exp{\imag\omega t} \phi(t)\,,&&
    \phi(t) = \int\dintbar{\omega} \exp{-\imag\omega t} \phi(\omega)\,,
    \elabel{Fourier}
\end{align}
and similarly for $\tildephi$,
with $\dbar\omega=\plaind\omega/(2\pi)$.

Our general approach in 
the three examples illustrated in Appendices \ref{appOU}, \ref{Sec:appendix-telegraphic-noise} and 
\ref{app-coupled-branchers}, where the extinction rate 
$sp_0$ is modulated by an external noise, is the following.
We first derive the action that governs the dynamics of the 
external noise $\mA_{\text{Y}}$ with Y either 
an Ornstein-Uhlenbeck process (Y$=$OU), a telegraphic noise (Y$=$T)
or a branching process (Y$=$BP).
Then, we derive the action $\mA_{\text{int}}$
that describes the interaction
between the noise Y and the branching process.
Merging the three parts, the action that encapsulates
all concurring processes is
\begin{equation}
    \mA = \mA_{\text{BP}} + \mA_{\text{Y}} + \mA_{\text{int}}\,.
\end{equation}
In each of the actions of the three subprocesses there are, or may
be, bilinear terms. We include those bilinear terms in the 
Gaussian model $\mA_0$ and 
group the rest in the perturbation $\mA_1$, so that
the overall action $\mA$ is written as
\begin{equation}
    \mA = \mA_0 + \mA_1\,.
\end{equation}
To calculate an observable, such as the expected number of 
infected individuals $\mean{N(t)}=\ave{\phi(t)\tildephi(t_0)}$,
we then perform a perturbative expansion about the Gaussian model,
\begin{equation}
\ave{\mO}=\ave{\int\mD\phi\mD\tildephi
\mD\psi\mD\tildepsi\ldots\,\, \mO \,\,
\exp{\mA_{1}[\phi,\tildephi,\psi,\tildepsi,\ldots]}}_0\,,
\end{equation}
where the fields $\psi,\tildepsi$ represent the external noise.


\section{Ornstein-Uhlenbeck Process}
\label{appOU}
\subsection{Field theory}
While some noisy processes can be described by a Doi-Peliti field theory (Sec.~\ref{Sec:TelegraphicNoise} and ~\ref{Sec:Coupled-Birth-Death-Process}), where fields capture the time-dependent density of a degree of freedom, others are easier described using a Langevin equation and the response field formalism \cite{MartinSiggiaRose:1973,Taeuber:2014}, where fields represent the degree of freedom itself. The Langevin equation of the Ornstein-Uhlenbeck process is\begin{align}\elabel{Langevin-OU}
    {\partial_t}y=-\beta y+\eta(t)
\end{align}
where $y(t)$ is the degree of freedom of the Ornstein-Uhlenbeck process, $\beta$ is the inverse persistence time, and $\eta(t)$ is a Gaussian white noise with mean $\mean{\eta}=0$, and correlator $\mean{\eta(t)\eta(t')}=2D\delta(t-t')$.

Using the Janssen-DeDominicis-Martin-Siggia-Rose (JDMSR) response field formalism \cite{MartinSiggiaRose:1973,DeDominicis:1976,Janssen:1976,Taeuber:2014}, the Langevin equation of the Ornstein-Uhlenbeck process, Eq.~\eref{Langevin-OU} can be transformed into the action \cite{Taeuber:2014}
\begin{align}\label{eq:OU_action}
    \mA_{\text{OU}}[\psi,\tildepsi]=\int\dint{t}\left\{
    -\tildepsi\left({\partial_t}+\beta\right)\psi+D\tildepsi^2
    \right\},
\end{align} 
which defines the field theory for the OU process. The field $\psi$ represents the values of the random variable in the OU process. The auxiliary field $\widetilde\psi$ is \emph{not} related to the $\psi$ field.
It is not to be confused with
 a Doi-shifted creator field. It is introduced in the JDMSR response field formalism purely to enforce the system to obey the OU Langevin equation, \eref{Langevin-OU}, \cite{Taeuber:2014}.

The bare propagator of the external noise is
\begin{equation}
    \tikz[baseline=-2.5pt,scale=0.75]{
\draw[Bactivity] (0:-0.8) -- (0:0.8);
}\corresponding
\ave{\psi(\omega)\tildepsi(\omega')}_{0} =
\frac{\deltabar(\omega+\omega')}{-\imag\omega+\beta} \,,
\end{equation}
and the coupling introduced by the constant $D$ 
in \eref{OU_action} is represented by
the source diagram
\begin{align}
\tikz[baseline=-2.5pt,scale=0.75]{
\draw[Bactivity] (0:0.8) -- (20:-0.8) node [at start, above] {$D$};
\draw[Bactivity] (0:0.8) -- (-20:-0.8);
}\,.
\end{align}

\label{sec:appendix-linear-OU}
\subsection{Mean-field approximation}\label{app_dawson} 
The average $\meanY{\mean{K(t)}}$ of $\mean{K(t)}$, Eq.~\eqref{eq_time_dependent_R_OU}, in the steady state of the Ornstein-Uhlenbeck process for $y$, Eq.~\eqref{eq_steady_state_OU}, requires us to determine the following integral
\begin{align}
  \meanY{\mean{K(t)}}=&s R_0  \int_{-\infty}^\infty\dint{y}\frac{1}{s+\lambda y}\sqrt{\frac{\beta}{2\pi D}}e^{-\frac{\beta y^2}{2D}}\nonumber\\
  =&\frac{s}{\lambda}\sqrt{\frac{\beta}{2\pi D}} R_0
  \int_{-\infty}^\infty\dint{x}\frac{e^{-x^2}}{\frac{s}{\lambda}\sqrt{\frac{\beta}{2D}}-x}\,,
  \end{align}
which contains the Hilbert transform of a Gaussian.
The Hilbert transform of a function $f$ is defined as 
\cite{Hilbert1912}\begin{align}
H[f(x)](z)=\frac{1}{\pi}\text{p.v.}\!\!\int_{-\infty}^\infty\dint{x}\frac{f(x)}{z-x}\,,
\end{align}
where p.v.~denotes Cauchy's principal value.
Using that the Dawson function \cite{Dawson1897}
\begin{align}
    \mD_+(z)=\exp{-z^2}\int_{0}^z\dint{t} \exp{t^2}.
    \elabel{Dawson_def}
\end{align}
is the Hilbert transform of a Gaussian,
\begin{equation}
    H\left[\exp{-x^2}\right]\left(z\right) = \frac{2}{\sqrt{\pi}} {\mD_+}\!\left(z\right) \,,
\end{equation}
we obtain the result in \eref{eq_R_0_OU}.

\subsection{Branching process coupled to an Ornstein-Uhlenbeck process}
\label{app:actionOU}
We couple the Ornstein-Uhlenbeck process in \eref{OU_action} to the
extinction rate of the branching process in \eref{birth-death-action}, as described by the master equation, 
\Eref{eq_master_branching_OU}.
From \eref{eq_master_branching_OU}, we derive the interaction term
\begin{align}
\label{eq:linear-OU-coupling}
\mA_{\text{int}}[\phi,\widetilde\phi,\psi]=
-\lambda\int\dint{t}\tildephi\phi\psi \,,
\end{align}
which produces the nonlinear coupling 
\begin{equation}
        \tikz[baseline=-2.5pt]{
\draw[Bactivity] (-30:0.8) -- (0,0) node [at end, above] {$-\lambda$};
\draw[Aactivity] (0:-0.8) -- (0:0.8);
}  \,.
\end{equation}
The overall action is then 
\begin{align}
    \mA=\mA_{\text{BP}}[\phi,\widetilde\phi]+\mA_{\text{OU}}[\psi,\tildepsi]+\mA_{\text{int}}[\phi,\widetilde\phi,\psi]\,,
    \elabel{action_lin_OU}
\end{align}
which combines 
the Doi-Peliti and JDMSR formalisms, drawing on \Erefs{all_def_r},~\eqref{eq:OU_action} and~\eqref{eq:linear-OU-coupling}. 
The subprocess governed by
the action $\mA_{\text{BP}}+\mA_{\text{int}}$ describes a branching processes with a factor $\psi$ 
that modulates 
 the extinction process. The expectation of any observable $\mathcal{O}$ of this subprocess is represented by a $n$-point correlation function $\ave{\mathcal{O}(\psi)}_\text{BP}$ that is a function of  $\psi$,
 \begin{align}
    \ave{\mathcal{O}(\psi)}_\text{BP}=\int\mathcal{D}[\phi,\widetilde\phi]\mathcal{O}e^{\mA_{\text{BP}}[\phi,\widetilde\phi]+\mA_\text{int}[\phi,\widetilde\phi,\psi]}.
\end{align}
The OU process renders $\psi$
a random variable with a probability
distribution given by (\ref{eq_steady_state_OU}). In order to take this distribution into account, the expectation $\ave{\mathcal{O}(\psi)}_\text{BP}$ needs to be considered as an observable for the OU process,
\begin{subequations}
\begin{align}
    \ave{\mathcal{O}(\psi)}=&\ave{\ave{\mathcal{O}}_\text{BP}}_\text{OU}\\
    =&\int\mathcal{D}[\psi,\widetilde\psi]\ave{\mathcal{O}(\psi)}_\text{BP}e^{\mA_{\text{OU}}[\psi,\widetilde\psi]}\\
    =&\int\mathcal{D}[\phi,\widetilde\phi,\psi,\widetilde\psi]\mathcal{O}e^{\mA[\phi,\widetilde\phi,\psi,\widetilde\psi]}.
\end{align}
\end{subequations}
In order to determine whether the dynamics are super- or subcritical, we calculate the expected number of infected individuals $\ave{N(t)}=\ave{\phi(t)\tildephi(0)}$ and look for its exponential growth and decay respectively. 
The extended action \eref{action_lin_OU}
introduces the following loop corrections to this propagator, 
\begin{subequations}
\label{eq:linearOU-OneLoopCorrections}
\begin{align}
\ave{\phi(t)\tildephi(0)}
\corresponding\,&
\tikz[baseline=-2.5pt]{
\draw[Aactivity] (0:-0.8) -- (0:0.8);
} +
\tikz[baseline=-2.5pt]{
\draw[Bactivity] (0.4,0.5) -- (-0.3,0);
\draw[Bactivity] (0.4,0.5) -- (0.1,0);
\draw[Aactivity] (0:-0.8) -- (0:0.8);
} \label{eq:linearOU-OneLoopCorrection}\\
&+\,\tikz[baseline=-2.5pt]{
\draw[Bactivity] (0.1,0.5) -- (-0.6,0);
\draw[Bactivity] (0.1,0.5) -- (-0.2,0);
\begin{scope}[xshift=0.5cm]
\draw[Bactivity] (0.3,0.5) -- (-0.4,0);
\draw[Bactivity] (0.3,0.5) -- (0.0,0);\end{scope}
\draw[Aactivity] (0:-0.8) -- (0:0.8);
} +
\tikz[baseline=-2.5pt]{
\draw[Bactivity] (0.1,0.5) -- (-0.6,0);
\draw[Bactivity] (0.1,0.5) -- (-0.3,0);
\begin{scope}[xshift=0.6cm]
\draw[Bactivity] (0.3,0.5) -- (-0.4,0);
\draw[Bactivity] (0.3,0.5) --  (-0.1,0);\end{scope}
\begin{scope}[xshift=0.2cm]
\draw[Bactivity] (0.3,0.5) -- (-0.4,0);
\draw[Bactivity] (0.3,0.5) -- (-0.1,0);\end{scope}
\draw[Aactivity] (0:-0.8) -- (0:0.8);
} +\dots\label{eq:linearOU-DysonSumCorrection}\\
&+\,\tikz[baseline=-2.5pt]{
\draw[Bactivity] (0.4,0.5) -- (-0.3,0);
\draw[Bactivity] (0.4,0.5) -- (0.1,0);;
\draw[Bactivity] (0.6,-0.5) -- (-0.5,0);
\draw[Bactivity] (0.6,-0.5) -- (0.3,0);
\draw[Aactivity] (0:-0.8) -- (0:0.8);
}
+\,\tikz[baseline=-2.5pt]{\begin{scope}[xshift=0.2cm]
\draw[Bactivity] (0.1,0.5) -- (-0.6,0);
\draw[Bactivity] (0.1,0.5) -- (-0.2,0);
\draw[Bactivity] (0.3,-0.5) -- (-0.4,0);
\draw[Bactivity] (0.3,-0.5) -- (0.0,0);\end{scope}
\draw[Aactivity] (0:-0.8) -- (0:0.8);
}+\dots\label{eq:linearOu-EntangledLoopCorrection}
\end{align}
\end{subequations}
The one-loop correction in \eref{linearOU-OneLoopCorrection} is
\begin{subequations}
\elabel{one_loop_correction}
\begin{align}
\tikz[baseline=-2.5pt]{
\draw[Bactivity] (0.4,0.5) -- (-0.3,0);
\draw[Bactivity] (0.4,0.5) -- (0.1,0);;
\draw[Aactivity] (0:-0.6) -- (0:0.6);
}
\corresponding&\int\dintbar{\omega}\dintbar{\omega'}\frac{2e^{-i\omega t}D\lambda^2}{(-i\omega+r)^2(-i(\omega+\omega')+r)(\omega'^2+\beta^2)} \nonumber\\
=&\int\dintbar{\omega}\frac{e^{-i\omega t}D\lambda^2}{(-i\omega+r)^2(-i\omega+r+\beta)\beta}\\
=&\frac{D\lambda^2}{\beta^3}e^{-rt}\Bigl(e^{-\beta t}-1+\beta t\Bigr)\,,
\end{align}
\end{subequations}
so that the propagator is
\begin{align}
    \ave{\phi(t)\tildephi(0)}= &
    \exp{-rt}\left(1+
    \frac{D\lambda^2}{\beta^3}e^{-rt}\Bigl(e^{-\beta t}-1+\beta t\Bigr)
    \right)\nonumber\\&
    +\mO\left(\left(\frac{D\lambda^2}{\beta^3}\right)^2\right)\,.
\end{align}
The pre-factor of $e^{-rt}$ indicates that, in this approximation, the critical point remains at $0=r=s(1-\mathbb{E}[K])$. Calculating only this one-loop approximation is indeed insufficient to see a shift in the critical point. 

Calculating, on the other hand, all loop corrections in \eref{linearOU-OneLoopCorrections} is doable in principle, but
calculating a closed form expression for each of them and doing
appropriate bookkeeping is an arduous task, in particular for 
entangled loops such as the ones in \eref{linearOu-EntangledLoopCorrection}.
Instead, we find that the Dyson sum, which includes all loop 
corrections of the form 
\eref{linearOU-OneLoopCorrection}-\eref{linearOU-DysonSumCorrection}, and excludes entangled loops, known as the non-approximation, provides a better approximation
to the expected number of infected individuals $\mean{N(t)}$.
This is,
\begin{subequations}\begin{align}
\mean{\Num(t)}_\text{Dy}
\corresponding\,&
\tikz[baseline=-2.5pt,scale=0.75]{
\draw[Aactivity] (0:-0.8) -- (0:0.8);
} +
\tikz[baseline=-2.5pt,scale=0.75]{
\draw[Bactivity] (0.4,0.5) -- (-0.3,0);
\draw[Bactivity] (0.4,0.5) -- (0.1,0);;
\draw[Aactivity] (0:-0.8) -- (0:0.8);
} 
+\,\tikz[baseline=-2.5pt,scale=0.75]{
\draw[Bactivity] (0.1,0.5) -- (-0.6,0);
\draw[Bactivity] (0.1,0.5) -- (-0.2,0);
\begin{scope}[xshift=0.5cm]
\draw[Bactivity] (0.3,0.5) -- (-0.4,0);
\draw[Bactivity] (0.3,0.5) -- (0.0,0);\end{scope}
\draw[Aactivity] (0:-0.8) -- (0:0.8);
} +
\dots\\
\corresponding&{\displaystyle\int}\frac{e^{-i\omega t}}{-i\omega+r}\sum\limits_{k=0}^\infty\left(\frac{D\lambda^2\dbar\omega}{\beta(-i\omega+r)(-i\omega+r+\beta)}\right)^k\elabel{geometric_sum}\\
    =&{\displaystyle\int}\frac{e^{-i\omega t}(-i\omega+r+\beta)\dbar\omega}{(-i\omega+r+\beta)(-i\omega+r)-\frac{D\lambda^2}{\beta}}\elabel{geometric_sum_partii}\\ 
    =&\,\Theta(t)e^{-(r+\frac{\beta}{2})t}
    \left(\cosh
    \left(\frac{\beta t}{2}\sqrt{1+\frac{4D\lambda^2}{\beta^3}}
    \right)
    \right.
    \nonumber\\&
    \left.
    +\frac{1}{\sqrt{1+\frac{4D\lambda^2}{\beta^3}}}\sinh
    \left(\frac{\beta t}{2}\sqrt{1+\frac{4D\lambda^2}{\beta^3}}
    \right)
    \right),
\elabel{TwoPolesSolution}
 \end{align}\end{subequations}
 where from \Eref{geometric_sum} to \Eref{geometric_sum_partii}, we calculated a geometric sum of the one-loop correction from \Eref{one_loop_correction}.
The exponential growth rate in \Eref{TwoPolesSolution}
gives the critical point approximated by the Dyson sum,
\begin{align}
    r+\frac{\beta}{2}\left(1-\sqrt{1+\frac{4D\lambda^2}{\beta^3}}\right)=0,
    \elabel{old_C15}
\end{align}
which forms a hypersurface in $r$-$\lambda$-$D$-$\beta$ space and which is transformed into the critical hypersurface in Eq.~\eqref{eq_critical_point_OU_noise}.




\section{Telegraphic noise}
\label{Sec:appendix-telegraphic-noise}
To cast the telegraphic noise in a field theory,
we consider two species, "on" and "off". 
The state of the telegraphic noise is given by the
number of particles $m$ and $n$ of species "on" and
"off" respectively.
Particles transmute between the two species 
according to \eref{tele_reactions} such that the
total number of particles is conserved.
In our case, the total number of particles is
$m+n=1$ since, initially, the telegraphic noise
is $T(t_0) = \eon$.

Using a \textit{bra-ket} notation, $\langle m,n|$ and $|m,n\rangle$, and ladder operators $a$, $a^\dagger$, $b$, and $b^\dagger$ with commutators $[a,a^\dagger]=1=[b,b^\dagger]$, $[a,b]=[a,b^\dagger]=[a^\dagger,b]=[a^\dagger,b^\dagger]=0$ and $a|m,n\rangle=m|m-1,n\rangle$, $a^\dagger|m,n\rangle=|m+1,n\rangle$, $b|m,n\rangle=n|m,n-1\rangle$, and $b^\dagger|m,n\rangle=|m,n+1\rangle$, the master equation \eqref{eq_master_telegraphic} can be turned into an equation for the probability generating function $|\mathcal{M}(t)\rangle:=\sum_{\{m,n\}}P(m,n;t)|m,n\rangle$ \cite{Doi:1976},
\begin{align}\label{eq:SQE-telegraphic-noise}
    {\partial_t}|\mathcal{M}(t)\rangle=\left(\mu_\textit{off}\left(b^\dagger-a^\dagger\right) a+\mu_\textit{on}\left(a^\dagger-b^\dagger\right) b\right)|\mathcal{M}(t)\rangle
\end{align}
Building on work by Peliti \cite{Peliti:1985}, Eq.~\eqref{eq:SQE-telegraphic-noise} can be turned into a field theory for fields $\psi$, $\widetilde\psi$ for $m$ particles 
in state "on", and $\chi$, $\tildechi$ for $n$ particles in state 
"off" with action $\mA_{T}$,
\begin{align}\label{eq:action-telegraphic-noise}
    \mA_{T}[\psi,\tildepsi,\chi,\tildechi]= \int
    \dint{t}
    & \left\{-
    \tildepsi(\partial_t+\moff)\psi
    -\tildechi(\partial_t+\mon)\chi \right.\nonumber\\&
    \left. +\mon \,\tildepsi\chi
    +\moff\, \tildechi\psi
    \right\}\,.
\end{align}

The values $\eon$ and $\eoff$, \Eref{telegraphic_noise_intro}, can be chosen arbitrarily. For simplicity, we set $\eoff=0$. 

To couple the telegraphic noise 
$T(t)$ to the branching process such that its value is added to the
death rate, we use the master equation from the previous model (branching with OU noise), \Eref{eq_master_branching_OU}, where we replace $\lambda y$ by $T(t)$. The value of the telegraphic noise $T(t)$ is represented by $\eon\psi^\dagger\psi$ in the field theory. The creator field is then Doi-shifted $\psi^\dagger=\widetilde\psi+1$, which leads to the following term in the action
 \begin{align}
    \mA_{\text{int}}[\phi,\widetilde\phi,\psi,\widetilde\psi]=
    -\eon\int\dint{t}\left\{\psi\widetilde\phi\phi+\widetilde\psi\psi\widetilde\phi\phi\right\}\,.
\end{align} 
As a result, the total extinction 
rate switches between values $sp_0$ and $sp_0+\eon$ in intervals that are exponentially distributed with rates $\mon$ and $\moff$, respectively.
The overall action of the process is then
\begin{align}\begin{split}
    \mA[\phi,\widetilde\phi,\psi,\widetilde\psi,\chi,\tildechi]=\hspace{-2cm}&\\
    &\mA_{\text{BP}}[\phi,\widetilde\phi]+\mA_{T}[\psi,\widetilde\psi,\chi,\tildechi]+\mA_{\text{int}}[\phi,\widetilde\phi,\psi,\widetilde\psi]\,.
\end{split}\end{align}
In this field theory we need the bare propagator
of the driving noise,
\begin{subequations}
\begin{align}
\tikz[baseline=-2.5pt,scale=0.75]{
\draw[Bactivity] (0:-0.8) -- (0:0.8);
\draw[black,fill= black] (0,0) circle (0.1);
} 
= &
\tikz[baseline=-2.5pt, scale=0.75]{
\draw[Bactivity] (0:-0.75) -- (0:0.75);
}
+
\tikz[baseline=-2.5pt, scale=0.85]{
\draw[Bactivity] (0:0.3) -- (0:0.75);
\draw[substrate] (0.3,0) -- (-0.3,0)  ;
\draw[Bactivity] (-0.3,0) -- (-0.75,0);
}
+
\tikz[baseline=-2.5pt, scale=0.85]{
\draw[Bactivity] (0.8,0) -- (0.55,0);
\draw[substrate] (0.55,0) -- (0.125,0)  ;
\draw[Bactivity] (0.125,0) -- (-0.125,0)  ;
\draw[substrate] (-0.125,0) -- (-0.55,0)  ;
\draw[Bactivity] (-0.55,0) -- (-0.8,0);
}
+\ldots \\
\corresponding&
\ave{\psi(\omega)\tildepsi(\omega')}_{0} \nonumber\\
=&
\frac{\deltabar(\omega+\omega')(-\imag\omega +\moff)}{(-\imag\omega+\mon)(-\imag\omega+\moff)-\mon\moff} \,,
\end{align}
\end{subequations}
and the two couplings
\begin{align}
    \tikz[baseline=-2.5pt]{
\draw[Bactivity] (-30:0.8) -- (0,0) node [at end, above] {$\eon$};
\draw[Aactivity] (0:-0.8) -- (0:0.8);
} 
&&\text{and}&&
    \tikz[baseline=-2.5pt]{
\draw[Bactivity] (210:0.8) -- (0,0) node [at end, above] {$\eon$};
\draw[Bactivity] (-30:0.8) -- (0,0) node [at end, above] {$\eon$};
\draw[Aactivity] (0:-0.8) -- (0:0.8);
} \,.
\end{align}

\noindent In order to identify the critical point, we calculate the expected number of individuals, 
\begin{subequations}
\begin{align}
    \mathbb{E}[N(t)]=&\langle\phi(t)\phi^\dagger(0)\psi^\dagger(0)\rangle\\
    =&\langle\phi(t)\widetilde\phi(0)\rangle+\langle\phi(t)\widetilde\phi(0)\widetilde\psi(0)\rangle \,.
    \elabel{tele_ave_N}
\end{align}
\end{subequations}
While the first term in \eref{tele_ave_N} is simply
\begin{align}\elabel{tele_prop_no_loops}
    \langle\phi(t)\widetilde\phi(0)\rangle=
    e^{-rt}\,,
\end{align}
the second term in \eref{tele_ave_N} is more complicated and can be represented in Feynman diagrams as a Dyson sum,
\begin{widetext}
\begin{subequations}
\begin{align}
    \langle\phi(t)\widetilde\phi(0)\widetilde\psi(0)\rangle
    \hat =&\,
    \tikz[baseline=-2.5pt]{
    \draw[Aactivity] (-0.5,0) -- (0.5,0);
    \draw[Bactivity] (0,0) -- (0.5,-0.25) node (A) [midway] {};
    \draw[black,fill= black] (A) circle (0.07);
    }
    +
    \tikz[baseline=-2.5pt]{
    \draw[Aactivity] (-0.5,0) -- (1.25,0);
    \draw[Bactivity] (0.75,0) -- (1.25,-0.25) node (A) [midway] {}; 
    \draw[Bactivity] (0.75,0) .. controls (0.75,-0.5) and (0,-0.5) .. (0,0) node (B) [midway] {};
    \draw[black,fill= black] (A) circle (0.07);
    \draw[black,fill= black] (B) circle (0.07);
    }
    +
    \tikz[baseline=-2.5pt]{
    \draw[Aactivity] (-0.5,0) -- (2,0);
    \draw[Bactivity] (1.5,0) -- (2,-0.25) node (A) [midway] {}; 
    \draw[Bactivity] (1.5,0) .. controls (1.5,-0.5) and (0.75,-0.5) .. (0.75,0) node (B) [midway] {}; 
    \draw[Bactivity] (0.75,0) .. controls (0.75,-0.5) and (-0.0,-0.5) .. (-0.0,0) node (C) [midway] {};
    \draw[black,fill= black] (A) circle (0.07);
    \draw[black,fill= black] (B) circle (0.07);
    \draw[black,fill= black] (C) circle (0.07);
    }
    +\cdots \\
    =&\int\frac{-\eon e^{-i\omega t}(-i\omega_1+\mon)}{(-i\omega+r)(-i(\omega-\omega_1)+r)((-i\omega_1+\moff)(-i\omega_1+\mon)-\moff\mon)} \nonumber\\
&\times\sum\limits_{k=0}^\infty\left(\int\frac{-\eon(-i\omega'+\mon)\dbar\omega'}{(-i(\omega-\omega')+r)((-i\omega'+\moff)(-i\omega'+\mon)-\moff\mon)}\right)^k\dbar\omega\dbar\omega_1\elabel{geometric_sum2}\\
=&\int\frac{-\eon e^{-i\omega t}(-i\omega+r+\mon)}{(-i\omega+r)\bigl((-i\omega+r+\moff)(-i\omega+r+\mon)-\moff\mon+\eon(-i\omega+r+\mon)\bigr)}\dbar\omega\elabel{poles-solution}\,,
\end{align}\end{subequations}
\end{widetext}

where from \Eref{geometric_sum2} to \Eref{poles-solution}, a geometric sum over loop corrections is calculated. Using the abbreviations,
\begin{subequations}
\begin{align}
    \epsilon =& \mon+\moff+\eon \,,\\
    \tau= & \sqrt{\epsilon^2-4\eon\mon} \,,
\end{align}
\end{subequations}
and based on \Erefs{tele_prop_no_loops} and~\eref{poles-solution}, the expected number of infected individuals can be calculated as\begin{align}
    \mathbb{E}[N(t)]=&\,\Theta(t)e^{-(r+\epsilon/2)t}\Bigl(\cosh(\tau t/2)\nonumber\\
    &\,+\frac{\mon+\moff-\eon}{\tau}\sinh(\tau t/2)\Bigr),
\end{align}

In fact, to determine the critical
point, we do not need to calculate
$\mathbb{E}[N(t)]$ explicitly because the boundary between supercritical and subcritical regimes is marked by a change of sign of the imaginary part of the complex poles of $\langle\phi(t)\widetilde\phi(0)\widetilde\psi(0)\rangle$ in Fourier space, \ie the poles of the $\omega$ integral in \Eref{poles-solution}. The equation for the critical hypersurface is then given by the equation that set the imaginary part of the $\omega$-poles equal to zero:
\begin{align}
\eon+\frac{r(r+\moff+\mon)}{r+\mon}=0,
\end{align}
which transforms into Eq.~\eqref{eq_tele_crit}, using \Eref{all_def_r_partii}.

\subsection{Correlation of the Telegraphic noise}\label{app-telegraphic-noise-correlation}
The Pearson correlation coefficient $\rho_{XY}$ of two random variables $X$ and $Y$ is defined as \cite{Bravais1844}\begin{align}
    \rho_{XY}=\frac{\mathbb{E}[XY]-\mathbb{E}[X]\mathbb{E}[Y]}{\sqrt{\mathbb{V}[X]\mathbb{V}[Y]}}
\end{align}
In the case of the telegraphic noise $T$, we are interested in $X=T(t)$ and $Y=T(t')$. If we assume that $t>t'>0$, then $\mathbb{E}[T(t)T(t')|T(0)=\eon]$ can be written in the field theory as\begin{align}
    \mathbb{E}[T(t)T(t')|T(0)=\eon]=\eon^2\ave{\psi(t)\psi^\dagger(t')\psi(t')\psi^\dagger(0)}.
\end{align}
Once the Doi-shift $\psi^\dagger=\tildepsi+1$ is performed, the only remaining non-vanishing term can be calculated as\begin{widetext}\begin{subequations}
\begin{align}
    \ave{\psi(t)\tildepsi(t')\psi(t')\tildepsi(0)}\hat =&\,\tikz[baseline=-2.5pt,scale=0.75]{
\draw[Bactivity] (0:-0.8) node[above] {$t$} -- (0:0.8);
\draw[black,fill= black] (0,0) circle (0.1);
}\tikz[baseline=-2.5pt,scale=0.75]{
\draw[Bactivity] (0:-0.8) node[above left] {$t'$} -- (0:0.8) node[above] {$0$};
\draw[black,fill= black] (0,0) circle (0.1);
} \\
    =&\,
\int\frac{e^{-i\omega (t-t')-i\omega't'}(-i\omega+\mon)(-i\omega'+\mon)\dbar\omega\dbar\omega'}{((-i\omega+\moff)(-i\omega+\mon)-\moff\mon)((-i\omega'+\moff)(-i\omega'+\mon)-\moff\mon)}\\
=&\,\frac{1}{(\moff+\mon)^2}\left(\mon+\moff e^{-(\moff+\mon)t'}\right)\left(\mon +\moff e^{-(\moff+\mon)(t-t')}\right).\elabel{telegraphic-noise-correlation-calc}
\end{align}\end{subequations} \end{widetext}
Changing the order of $t$ and $t'$ such that $t'>t$ amounts to swapping each $t$ for a $t'$ and vice versa in \Eref{telegraphic-noise-correlation-calc}. In the steady state, only the difference between $t$ and $t'$ enters and a single observation of $T$ will be a Bernoulli experiment, in which $T=\eon$ is drawn with probability $\mon/(\mon+\moff)$. Hence its expectation equals $\mathbb{E}[T]=\eon\mon/(\mon+\moff)$ and its variance equals $\mathbb{V}[T]=\eon^2\mon\moff/(\mon+\moff)^2$, consistent with the second moment \Eref{telegraphic-noise-correlation-calc} for $t-t'\rightarrow\infty$ and $t'\rightarrow\infty$.  Combining \Eref{telegraphic-noise-correlation-calc} with the exepectation and variance of $T$, we find the Pearson correlation coefficient in \Eref{telegraphic-noise-correlation}.

\section{Two coupled branching process}
\label{app-coupled-branchers}
The two populations of branching species A and B are represented by fields 
$\phi$, $\tildephi$ and $\psi$, $\tildepsi$ respectively,
and both follow the dynamics in the branching action 
\eref{birth-death-action} with coefficients $r_A$, $q_{jA}$
and $r_B$, $q_{jB}$, respectively. The action of the coupled branching 
process then includes the branching actions $\mA_{\text{BP}}[\phi,\tildephi]$ and $\mA_{\text{BP}}[\psi,\tildepsi]$, plus
a third term that describes the interactions 
in \eref{AB_interaction} between the
two species,
\begin{align}
\mA_{\text{int}}[\phi,\widetilde\phi,\psi,\widetilde\psi]=\int
\dint{t}\left\{\muA(\widetilde\psi-\widetilde\phi)\phi+\muB(\widetilde\phi-\widetilde\psi)\psi\right\}\,.
\end{align}
This introduces additional mass terms, so that the bare propagators read \begin{subequations}
\begin{align}
    \tikz[=-2.5pt, scale=0.75]{\draw[Aactivity] (0:0) -- (0:0.8);}\,\corresponding&\,\frac{\deltabar(\omega+\omega')}{-i\omega+r_A+\muA},\\
    \tikz[=-2.5pt, scale=0.75]{\draw[Bactivity] (0:0) -- (0:0.8);}\,\corresponding&\,\frac{\deltabar(\omega+\omega')}{-i\omega+r_B+\muB},
\end{align}\end{subequations}
as well as the interaction vertices
\begin{align}
    \tikz[baseline=-2.5pt, scale=0.75]{
\draw[Aactivity] (0:0) -- (0:-0.8) node [at start, above] {$\muB$};
\draw[Bactivity] (0:0.8) -- (0:0);
}
&&\text{and}&&
\tikz[baseline=-2.5pt, scale=0.75]{
\draw[Aactivity] (0:0) -- (0:0.8) node [at start, above] {$\muA$};
\draw[Bactivity] (0:-0.8) -- (0:0);
}
\end{align}
The overall action of the coupled branching processes is then
\begin{align}
    \mA[\phi,\widetilde\phi,\psi,\widetilde\psi]=\mA_{\text{BP}}[\phi,\widetilde\phi]+\mA_{\text{BP}}[\psi,\widetilde\psi]+\mA_{\text{int}}[\phi,\widetilde\phi,\psi,\widetilde\psi].
\end{align}
Since the interaction terms in the action are all bilinear,
we can include them in the Gaussian model
$\mA_0$. Then, to find the critical point, all
we need are the propagators,
\begin{subequations}\elabel{propABBA}
\begin{align}
    \tikz[baseline=-2.5pt,scale=0.75]{
\draw[Aactivity] (0:-0.8) -- (0:0.8);
\draw[black,fill= black] (0,0) circle (0.1);
} 
= &
\tikz[baseline=-2.5pt, scale=0.75]{
\draw[Aactivity] (0:-0.75) -- (0:0.75);
}
+
\tikz[baseline=-2.5pt, scale=0.75]{
\draw[Aactivity] (0:0.3) -- (0:0.75);
\draw[Bactivity] (0.3,0) -- (-0.3,0)  ;
\draw[Aactivity] (-0.3,0) -- (-0.75,0);
}
+
\tikz[baseline=-2.5pt, scale=0.75]{
\draw[Aactivity] (0.8,0) -- (0.55,0);
\draw[Bactivity] (0.55,0) -- (0.125,0)  ;
\draw[Aactivity] (0.125,0) -- (-0.125,0)  ;
\draw[Bactivity] (-0.125,0) -- (-0.55,0)  ;
\draw[Aactivity] (-0.55,0) -- (-0.8,0);
}
+\ldots 
\elabel{prop_AA}\\
\corresponding &
\ave{\phi(\omega)\tildephi(\omega')}_0 \\
=&
\frac{\deltabar(\omega+\omega')(\imag\omega+r_B+\muB)}{(-\imag\omega+r_A+\muA)(-\imag\omega+r_B+\muB)-\muA\muB} \,,\\
\tikz[baseline=-2.5pt, scale=0.75]{
\draw[Aactivity] (0:0) -- (0:0.8);
\draw[Bactivity] (0:-0.8) -- (0:0);
\draw[black,fill= black] (0,0) circle (0.1);
}
= &
\tikz[baseline=-2.5pt, scale=0.75]{
\draw[Aactivity] (0:0) -- (0:0.8);
\draw[Bactivity] (0:-0.8) -- (0:0);
}
+
\tikz[baseline=-2.5pt, scale=0.75]{
\draw[Aactivity] (0.8,0) -- (0.45,0);
\draw[Bactivity] (0.45,0) -- (0,0) ;
\draw[Aactivity] (0,0) -- (-0.35,0) ;
\draw[Bactivity] (-0.35,0) -- (-0.8,0);
}
+
\tikz[baseline=-2.5pt, scale=0.75]{
\draw[Aactivity] (0.9,0) -- (0.65,0);
\draw[Bactivity] (0.65,0) -- (0.3,0)  ;
\draw[Aactivity] (0.3,0) -- (0.05,0)  ;
\draw[Bactivity] (0.05,0) -- (-0.3,0)  ;
\draw[Aactivity] (-0.3,0) -- (-0.55,0) ;
\draw[Bactivity] (-0.55,0) -- (-0.9,0);
}
+\ldots \,,\\
\corresponding &
\ave{\psi(\omega)\tildephi(\omega')}_0 \\
=&
\frac{\deltabar(\omega+\omega')\muA}{(-\imag\omega+r_A+\muA)(-\imag\omega+r_B+\muB)-\muA\muB} \,.
\elabel{prop_BA}
\end{align}
\end{subequations}
The first moments of the particle numbers are then derived using inverse Fourier transforms:
\begin{subequations}
\elabel{exp_N_BP_BP}
\begin{align}
\mathbb{E}[N_A(t)&|N_A(0)=1,N_B(0)=0]=\ave{\phi(t)\widetilde\phi(0)}_0
\nonumber\\
=&\,\frac{e^{-\epsilon t}}{\tau}\left((r_B+\mu_B-\epsilon)\sinh(\tau t)+\tau\cosh(\tau t)\right)\\
\mathbb{E}[N_A(t)&|N_A(0)=0,N_B(0)=1]=\ave{\phi(t)\widetilde\psi(0)}_0
\nonumber\\
=&\,\frac{\mu_A e^{-\epsilon t}}{\tau}\sinh(\tau t),
\end{align}
\end{subequations}
where
\begin{subequations}
\begin{align}
    \epsilon=&\,\frac{r_A+r_B+\muA+\muB}{2}\\
    \tau=&\,\frac{\sqrt{(r_A-r_B+\muA-\muB)^2+4\muA\muB}}{2}
\end{align}
\end{subequations}

The propagators, \Eref{propABBA}, readily encode the critical
point of the coupled branching process. Both contour integrals have two poles,
\begin{align}
    \omega=&-\imag\left(\epsilon\pm\tau\right)\,,
\end{align}
which are purely imaginary. In the subcritical regime, their imaginary part is negative, while in the supercritcal regime at least one of the poles has a positive imaginary part. Thus the critical hypersurface is determined as
\begin{equation}
    r_Ar_B+r_A\muB+r_B\muA =0 \,,
\end{equation}
which is transformed into \eref{critical_line_coupled_branchers}.

\bibliography{books,articles}
\end{document}